\documentclass[times,twocolumn,review,nopreprintline]{elsarticle}
\usepackage{arxiv}

\usepackage{framed,multirow}
\usepackage{algorithmic}

\usepackage{latexsym}

\usepackage{url}
\usepackage{xcolor}
\usepackage{array}

\usepackage{hyperref}
\hypersetup{%hypertex=true,
	colorlinks=true,
	linkcolor=red,
	anchorcolor=green,
	citecolor=green}
\usepackage[footnotehyper]{nicematrix}

\definecolor{newcolor}{rgb}{.8,.349,.1}

\usepackage{bm}
\usepackage{color}
\usepackage{multirow}
\usepackage{bbding}
\definecolor{Emerald}{rgb}{0.31, 0.78, 0.47}
\definecolor{Denim}{rgb}{0.08, 0.38, 0.74}
\definecolor{Green(ryb)}{rgb}{0.4, 0.69, 0.2}
\definecolor{Iris}{rgb}{0.35, 0.31, 0.81}

\usepackage{amsmath,amssymb,amsfonts}
\usepackage{graphicx}
\usepackage{textcomp}
\usepackage{threeparttable}
\usepackage{subfigure}
\usepackage{caption}
\usepackage{booktabs}
\usepackage{arydshln}
\usepackage{supertabular}
\usepackage{float}
\usepackage[ruled,vlined]{algorithm2e}

\begin{document}

\verso{Yutong Xie \textit{et~al.}}

% \begin{frontmatter}

\title{Attention Mechanisms in Medical Image Segmentation: A Survey}

\author[1]{Yutong \snm{Xie}\fnref{fn1}}
\author[2]{Bing \snm{Yang}\fnref{fn1}}
\fntext[fn1]{Co-first authors; yutong.xie678@gmail.com; yangbing@mail.nwpu.edu.cn}
\author[2]{Qingbiao \snm{Guan}}
\author[3]{Jianpeng \snm{Zhang}}
\author[1]{Qi \snm{Wu}\corref{cor1}}
\author[2]{Yong \snm{Xia}\corref{cor1}}
\cortext[cor1]{Corresponding authors; yxia@nwpu.edu.cn; qi.wu01@adelaide.edu.au}

\address[1]{University of Adelaide, Adelaide, Australia}
\address[2]{Northwestern Polytechnical University, Xi'an, China}
\address[3]{Alibaba DAMO Academy, China}

% \received{1 May 2013}
% \finalform{10 May 2013}
% \accepted{13 May 2013}
% \availableonline{15 May 2013}
% \communicated{S. Sarkar}

\begin{abstract}
Medical image segmentation plays an important role in computer-aided diagnosis. Attention mechanisms that distinguish important parts from irrelevant parts have been widely used in medical image segmentation tasks. 
This paper systematically reviews the basic principles of attention mechanisms and their applications in medical image segmentation. 
First, we review the basic concepts of attention mechanism and formulation. 
Second, we surveyed over 300 articles related to medical image segmentation, and divided them into two groups based on their attention mechanisms, non-Transformer attention and Transformer attention. 
In each group, we deeply analyze the attention mechanisms from three aspects based on the current literature work, \textit{i.e.}, the principle of the mechanism (what to use), implementation methods (how to use), and application tasks (where to use). 
We also thoroughly analyzed the advantages and limitations of their applications to different tasks. 
Finally, we summarize the current state of research and shortcomings in the field, and discuss the potential challenges in the future, including task specificity, robustness, standard evaluation, \textit{etc}. 
We hope that this review can showcase the overall research context of traditional and Transformer attention methods, provide a clear reference for subsequent research, and inspire more advanced attention research, not only in medical image segmentation, but also in other image analysis scenarios.
\end{abstract}


% \end{frontmatter}

\maketitle

%\linenumbers

%% main text
\section{Introduction}
Medical image segmentation, as an important and difficult part of computer-aided diagnosis (CAD), has attracted much attention in recent studies. Its purpose is to differentiate anatomical or pathological structures in various medical images, such as computed tomography (CT), magnetic resonance imaging (MRI), positron emission tomography (PET), X-ray, ultrasound imaging (UI), and common RGB images like microscopy and fundus retinal images. Accurate segmentation of medical images is advantageous for diagnosis, treatment, and prognosis. The automation of this task, however, is extremely challenging due to three reasons: (1) the low soft tissue contrast results in fuzzy object boundaries; (2) anatomical or pathological structures may vary greatly in shape, size, and location; and (3) it is difficult to obtain sufficient annotated medical images to train segmentation models constrained by the labor cost and expertise. This makes it difficult to model the semantic relationship between different objects and backgrounds properly. 

\begin{figure}[!h]
    \centering
    \centerline{\includegraphics[width=\linewidth]{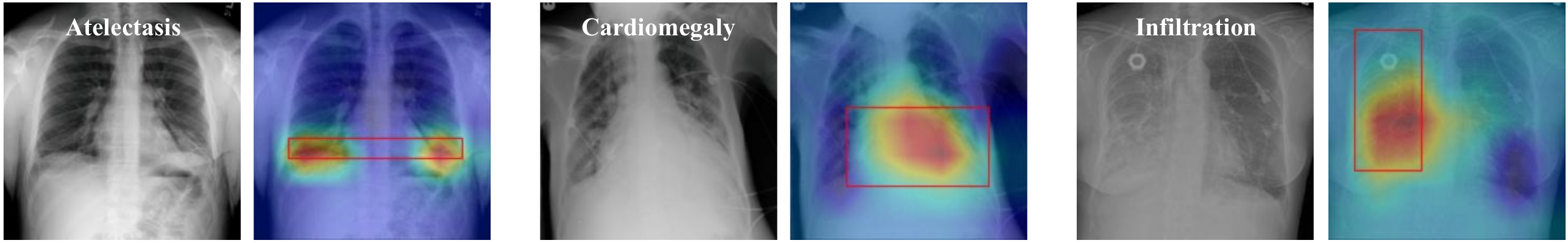}}
    \caption{Visualization of three chest X-ray samples with their attention maps obtained by~\cite{hu2018squeeze}. The red rectangles indicate the ground truth bounding box including disease regions.}
    \label{vis}
\end{figure}

In the human visual cognition system, we are naturally skilled at focusing on the area of interest and ignoring the interference of other background information, which helps us to identify and judge more accurately and efficiently. 
Imitating this, attention mechanisms are proposed to adaptively assign weights to different regions in an image, enabling neural networks to focus on the important regions related to the target task and disregard irrelevant areas, as shown in Figure~\ref{vis}. 
This powerful capability is well-suited for capturing complex semantic relationships in medical image segmentation. Furthermore, the attention mechanism can be utilized to explain the correlation between input and output data, illustrating what the model has learned, thus providing us with an interpretability insight into the black box of neural networks.

\begin{figure}[!h]
    \centering
    \centerline{\includegraphics[width=\linewidth]{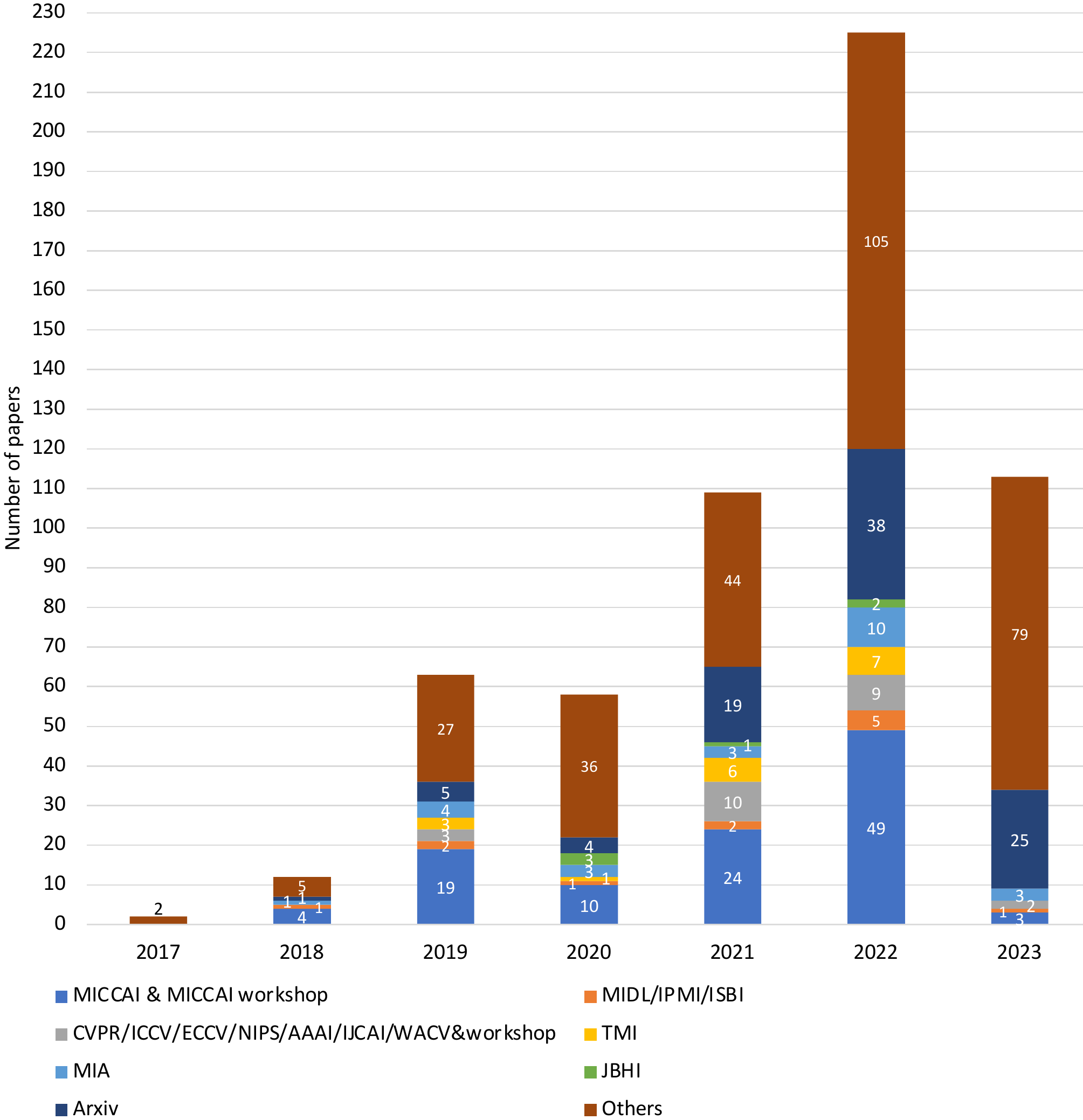}}
    \caption{Number of medical image segmentation papers from 2017 to 2023 whose titles include the word "Attention" or "Transformer". Different colors represent different sources of selected papers.}
    \label{Trend}
\end{figure}

Over the past few years, attention mechanisms have played an increasingly important role in medical image segmentation. Figure~\ref{Trend} illustrates that the number of papers focused on attention mechanisms in medical image segmentation has seen a steady annual growth from 2017 to 2023 and lists the source of published papers. 
Convolutional neural networks (CNNs) and their variants have shown advantages in medical image segmentation tasks due to their ability to learn hierarchical features. Attention mechanisms were first introduced as a plug-in to CNNs, allowing the network to adjust weights dynamically based on features. For example, works in~\cite{wang2019abdominal,oktay2018attention} improved the segmentation performance of abdominal organs by integrating the attention mechanism into U-Net.
To further model long-distance dependencies, the attention mechanism of the Transformer was proposed(~\cite{vaswani2017attention}), as demonstrated in the natural language processing field. Since the emergence of Vision Transformer in 2020, the Transformer attention mechanism has made significant breakthroughs in many visual tasks, drawing the attention of the medical community. As a result, there has been a rapid increase in Transformer-based papers related to medical image segmentation.

To keep up with the rapid increase in attention-based medical segmentation research, a survey of existing relevant works is urgently needed to provide the latest and comprehensive view of new research.
The purpose of this paper is to summarize and categorize the current attention-based methods of medical image segmentation, while offering thoughtful commentary on the current state of the field and making suggestions for future research.
Here, we classify the research into two types: traditional attention (Non-Transformer-based in this survey) and Transformer-based attention, as Transformer-based methods are increasingly recognized as a distinct and mainstream category in medical image segmentation research. Notably, Transformer blocks can serve as basic network blocks, while traditional attention should be combined with convolutional layers to form a block, which is also one of the bases for our classification.

In addition, we briefly compare this paper to various existing surveys which have reviewed attention methods and Transformer in medical image analysis.~\cite{shamshad2023transformers},~\cite{he2022transformers},~\cite{azad2023advances} and~\cite{li2023transforming} provided a survey of Transformer-based applications in medical image analysis. In contrast, our work reviews the entire domain of attention mechanisms beyond just Transformer-based methods.~\cite{gonccalves2022survey} surveyed attention mechanisms in medical image analysis tasks. However, segmentation is a challenging and critical task in medical image analysis, and the number of methods for this task exceeds that of the other tasks combined. Therefore, we have focused our summary of attention-based methods specifically on medical image segmentation on providing a deeper, task-specific understanding.

\begin{figure}[!h]
    \centering
    \centerline{\includegraphics[width=\linewidth]{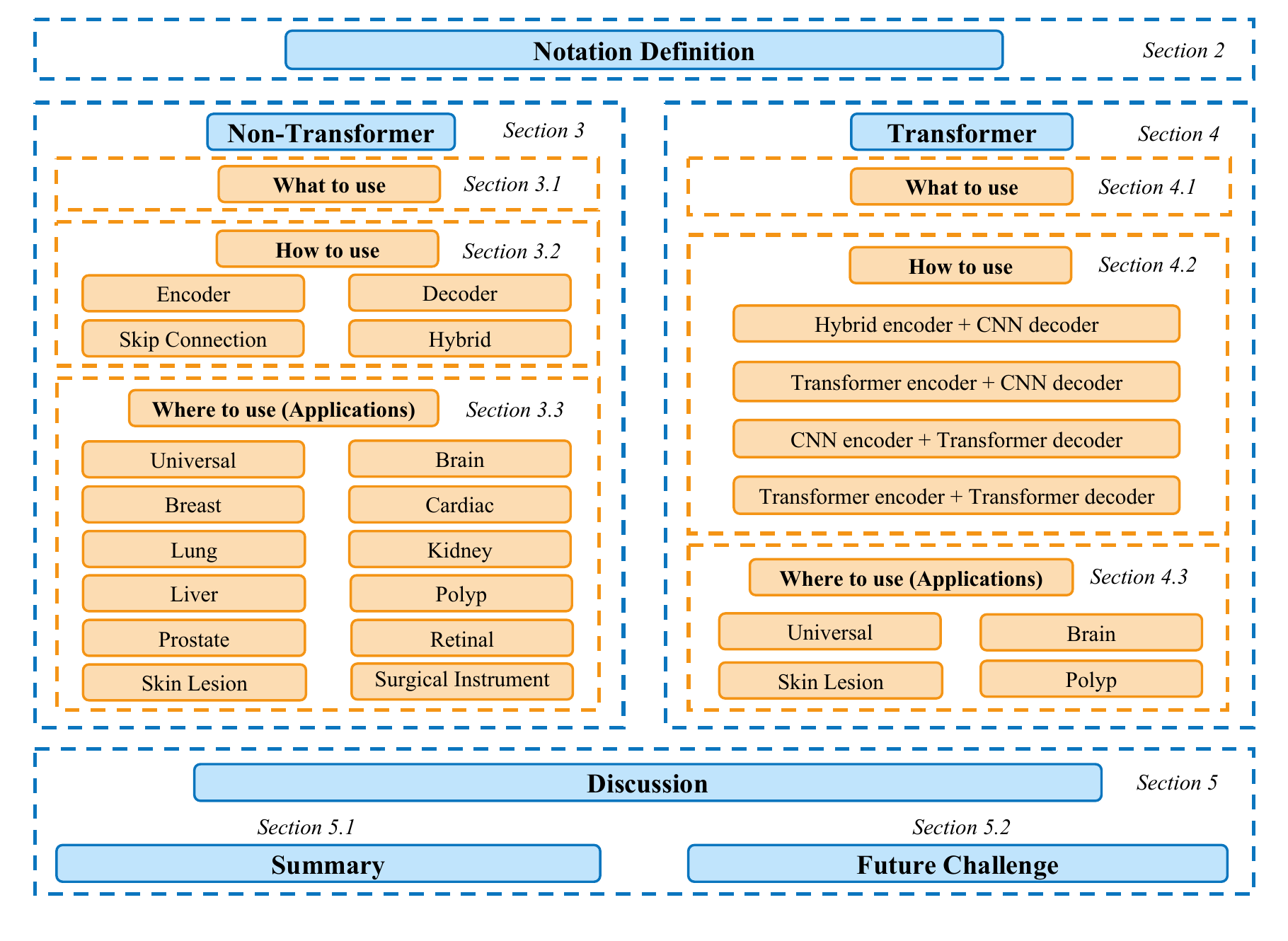}}
    \caption{Schematic structure of attention mechanisms in the medical segmentation and the relationship between the adjacent sections. The body of this survey mainly contains the notation definitions, Non-Transformer applications, Transformer applications, discussions, and future challenges.}
    \label{Transformer}
\end{figure}

The rest of the paper is organized as follows (see Figure~\ref{Transformer}). Section~\ref{sec2} is the notation definition. Section~\ref{sec3} and Section~\ref{sec4} are the main parts of our survey, in which we revisit the basic principles of attention mechanism separately and extensively review the existing Non-Transformer-based and Transformer-based medical segmentation methods following our taxonomies. 
We summarize the main conclusion of this survey in and highlight several future challenges in the study of attention-based algorithms for medical image segmentation tasks in Section~\ref{sec5}.

%在这一部分增加对下面表格和数据集的说明，对于每个应用指出一些常用的数据集（仅列举，不再做具体说明），并说明选取的常用性能指标及对应的缩写
\section{Notation Definition}
\label{sec2}
Table~\ref{table1} gives some notation definitions mentioned in the following chapters to aid understanding.

\renewcommand\arraystretch{1.25}
\begin{table}[h]
\caption{Key notations in this paper.}
\resizebox{\linewidth}{!}{
\begin{tabular}{@{}cllll@{}}
\toprule
\multicolumn{1}{l}{Type}    & Notation & Meaning                                                                   &  &  \\ \midrule
                            & NM       & Not mentioned                                                            &  &  \\
                            & NaN      & Not a number                                                             &  &  \\
                            & A+B      & Join the two datasets A and B together to train or test                                                            &  &  \\
                            & Enc      & Encoder                                                             &  &  \\
                            & Dec      & Decoder                                                             &  &  \\
\multirow{-6}{*}{Universal} & Skip.      & Skip connection                  &  &  \\ \midrule
                            & P        & Precision                                                                &  &  \\
                            & F1       & F1-score                                                                 &  &  \\
                            & HD       & Hausdorff distance                                                       &  &  \\
                            & JC       & Jaccard coefficent                                                       &  &  \\
                            & JI       & Jaccard index                                                            &  &  \\
                            & OE       & Overlapping error                                                       &  &  \\
                            & SE       & Sensitivity                                                              &  &  \\
                            & SP       & Specificity                                                              &  &  \\
                            & ACC      & Accuracy                                                                 &  &  \\
                            & AUC      & Area under receiver operating characteristic (ROC) &  &  \\
                            & FDR      & False discovery rate                                                     &  &  \\
                            & MSD      & Mean surface distance                                                    &  &  \\
                            & OE      & Overlapping error                                                   &  &  \\
\multirow{-12}{*}{Metric}   & mIoU     & Mean intersection-over-union                                             &  &  \\ \bottomrule
\end{tabular}
}
\label{table1}
\end{table}

\section{Non-Transformer in Medical Segmentation}
\label{sec3}

To provide researchers with a more comprehensive and clear understanding of the applications of Non-Transformer attention mechanisms in medical image segmentation, we will progressively categorize these studies into three aspects:
(1) the types of attention mechanisms used in medical image segmentation (what to use),
(2) the locations in the network where attention mechanisms are utilized (how to use), and
(3) the specific clinical tasks where attention mechanisms are applied (where to use).

\subsection {What to Use}

\subsubsection{Definition}
%该段视情况考虑是否还需要增加各种attention类型下的具体实例
The attention mechanism is inspired by the recognition process of the human visual system, allowing networks to focus on specific objects while ignoring irrelevant areas. This provides localized classification information, which is often desirable in image-processing networks. We will briefly introduce the theoretical basis of the attention mechanism and then illustrate common attention mechanisms in medical image segmentation tasks.

Following the description and taxonomy by~\cite{guo2022attention}, almost all
existing attention mechanisms can be formulated as
\begin{equation}
    {\rm Attention} = f(g(x), x)
\end{equation}
where $g(x)$ is a specific kind of generated attention. $f(g(x), x)$ represents processing input vector $x$ based on the attention $g(x)$ which is consistent with processing critical regions and getting information. 
Take self-attention as an example, we first transformed $x$ into three distinct matrix representations: queries $Q \in \mathbb{R}^{n\times d_{q}}$, keys $K \in \mathbb{R}^{n\times d_{k}}$, and values $V \in \mathbb{R}^{n\times d_{v}}$, all with dimensions $d_{q}=d_{k}=d_{v}=d_{model}$. $g(x)$ and $f(g(x), x)$ can be represented as
\begin{equation}
\begin{split}
    g(x) = {\rm softmax}(\frac{QK^{T}}{\sqrt{d_{k}}}) \\
    f(g(x), x)=g(x)V
\end{split}
\end{equation}
where $QK^T$ computes the relevance score between different entities, $d_{k}$ is the scaling factor, softmax operation translates the score into probability and multiplying with $V$ is to obtain the weighted matrix. 

\subsubsection{Non-Transformer attention}
\begin{figure}[!h]
    \centering
    \centerline{\includegraphics[width=\linewidth]{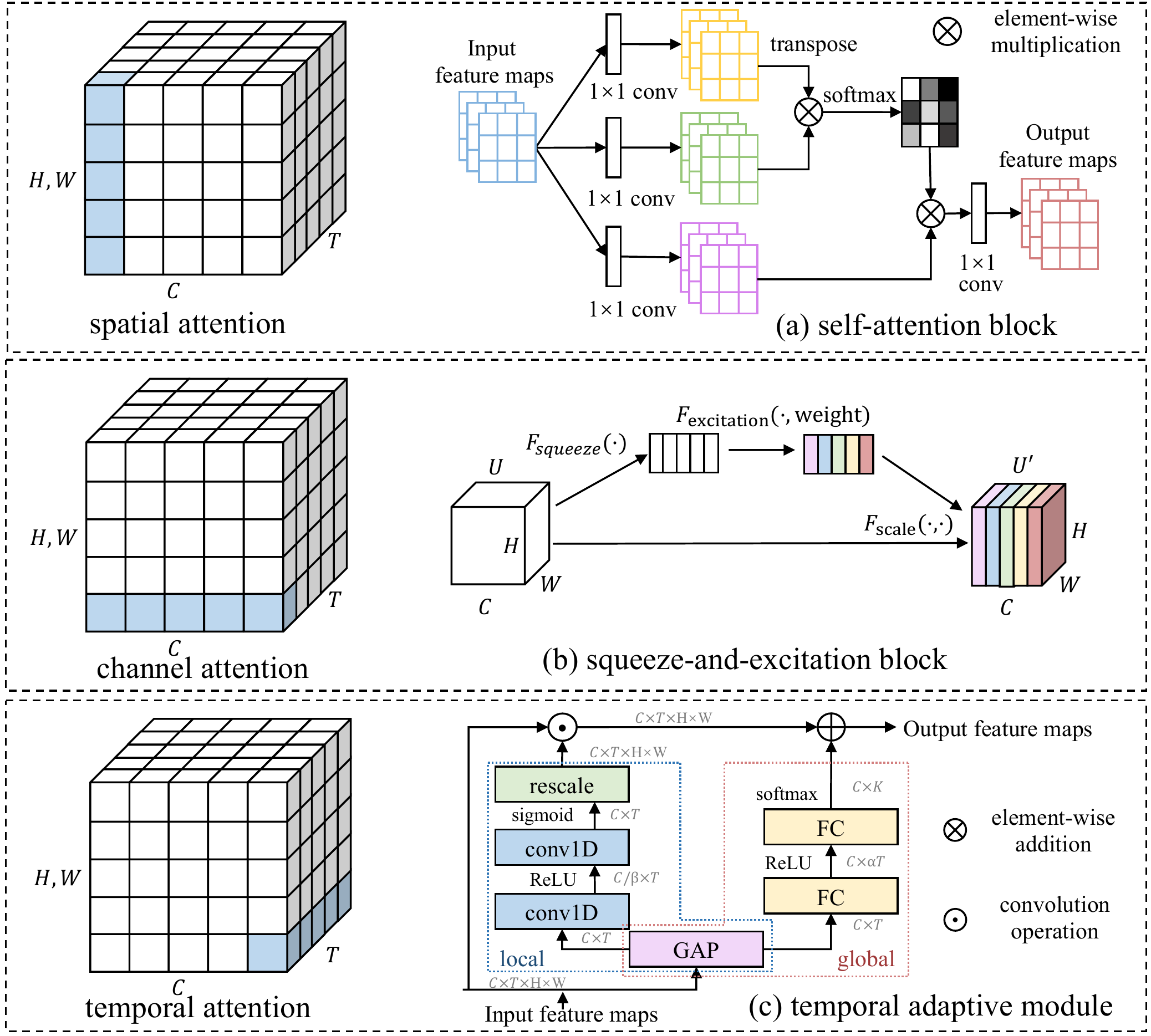}}
    \caption{Spatial, channel and temporal attentions, which follow~\cite{guo2022attention}. Figure (a)(b)(c) follow~\cite{zhang2019self},~\cite{hu2018squeeze} and~\cite{liu2021tam}.}
    \label{att}
\end{figure}

Commonly used Non-Transformer attention mechanisms can be categorized into three main types: channel, spatial, and temporal attention, as shown in Figure~\ref{att}. These attention mechanisms can be used individually or in combination to enhance the network's performance in medical image segmentation tasks.

\textbf{Channel attention} adaptively reweights each channel, treating them as representing different objects~(\cite{chen2017sca}). The concept of channel attention was first introduced in squeeze-and-excitation networks (SENet) by~\cite{hu2018squeeze}, using a squeeze-and-excitation (SE) block to capture channel-wise relationships. The squeeze module converts each channel into a single value using global average pooling, while the excitation module outputs an attention vector through fully-connected and non-linear layers. Subsequent works aimed to improve the squeeze or excitation process. For example,~\cite{qin2021fcanet} treated global average pooling as a case of the discrete cosine transform in the squeeze module and proposed the frequency channel attention net (FcaNet) based on information compression.~\cite{Wang_2020_CVPR} proposed an efficient channel attention (ECA) block, which takes the direct interaction between each channel and its k-nearest neighbors into account to improve the excitation module with less complexity. Meanwhile,~\cite{lee2019srm} adopted style pooling in the squeeze process and inserted a channel-wise fully-connected layer in the excitation module to reduce the computational cost in both steps.

In the medical image segmentation, SE block~(\cite{hu2018squeeze}) has been applied by~\cite{yang2021improved}. To avoid the loss of information in the SE block,~\cite{xie2021duda} proposed introducing multiscale context information via parallel dilated convolution with different dilation rates.
Other channel attention methods~\cite{wang2021hybrid,wang2019dual,guo2021channel} have also been designed for specific tasks, focusing on channels with ample information while restraining irrelevant channels.

\textbf{Spatial attention} helps to identify the important regions in an image by assigning importance scores to different spatial regions in the feature map (bounded by width and height).
% The spatial transformer network (STN)~\cite{jaderberg2015spatial} predicts relevant regions explicitly by giving an affine transformation while learning geometric transformation invariance. Some may implicitly indicate the important regions through element-wise production or aggregating information from an attention map. 
Attention gates~(\cite{oktay2018attention}) utilize additive attention between the input and a gate signal collected at a coarse scale to obtain the gating coefficient for building a spatial attention weight map.  It provides a modular and light paradigm highlighting important areas and suppressing features in unrelated regions. 
GENet~(\cite{hu2018gather}) introduces a spatial recalibration function called a gather-excite module, similar to SENet~(\cite{hu2018squeeze}), and then uses interpolation to form an attention map. 
To increase the receptive field, self-attention has also been introduced. 
~\cite{wang2018non} proposed the Non-local network, which augments each pixel of the convolutional features with contextual information (the weighted sum of the whole feature map) to encode the correlated patches in a long-range fashion.

We observe that attention gates~(\cite{oktay2018attention}), and Non-local networks~(\cite{wang2018non}) often appear in medical image segmentation.~\cite{oktay2018attention} proposed the attention gate to guide the model’s attention on targeted regions with a gating signal collecting from a course scale for abdominal multi-organ segmentation, and the attention gate is widely used and adapted in medical image segmentation~(\cite{duran2022prostattention,kearney2019attention}). While in the attention map comes from the resampling of the prostate prediction instead of the preceding convolutional block in~\cite{duran2022prostattention}. And~\cite{kearney2019attention} substitute the spatial attention coefficients for the global gating signal as opposed to previous study~(\cite{oktay2018attention}) to obtain more sensitive spatial information.

Besides,~\cite{mostayed2019content} applied self-attention multiplication with decoder feature maps to reduce the number of parameters and learn the object boundaries.~\cite{ding2020high} proposed high-order attention, which allows each pixel to build its own global attention map and then constructs the attention map through graph transduction, thus capturing relevant context information at high order to enhance relevant pixels.~\cite{xu2021exploiting} proposed a vector self-attention layer for long-range spatial reasoning with geometric priors and multi-scale calibration. 

Furthermore, some customized spatial attentions are designed to solve specific problems, $e.g.$, blur boundaries. Some may develop the edge attention~(\cite{zhang2019net,wang2020aec}), while others regard the shape as prior knowledge to guide a shape attention~(\cite{li2020joint,zhang2021dense}).

\textbf{Temporal attention} is usually seen as a dynamic time frame selection mechanism when the data has a temporal dimension, $e.g.$, a video.~\cite{li2019global} propose the global-local representation (GLTR) for temporal relation modeling through self-attention. While~\cite{liu2021tam} propose the temporal adaptive module (TAM), adopting an adaptive kernel instead of self-attention with a local branch and global branch to capture complex temporal relationships with lower computational costs.
Temporal attention may be applied in multi-frame images or videos clinically.~\cite{jin2019incorporating} utilize the inherent temporal clues from the instrument motion as prior.~\cite{ahn2021multi} adopt spatio-temporal attention in multi-time frames to obtain interframe consistency among the images.

Besides, spatial attention may combine with channel or temporal attention to obtain more comprehensive information. The convolutional block attention module (CBAM,~\cite{woo2018cbam}) calculates channel and spatial attention in serial separately, in which the channel attention module utilizes two parallel branches via max-pool, and avg-pool operations and the spatial attention module applies a convolution layer with a larger kernel to generate the attention map. Thus, CBAM can emphasize useful channels and enhance informative local regions. The dual attention network (DANet, \cite{fu2019dual}) also computes channel and spatial attention separately via self-attention and fuses them for final results. Moreover, the spatial attention in CBAM~(\cite{woo2018cbam})(with max-pooling and average-pooling) is also introduced with the original version~(\cite{guo2021sa}) and improved version(\emph{e.g.,}~\cite{guo2020residual,jiang2021bi}), in which~\cite{guo2020residual} apply max-pooling and average-pooling operations along the
channel axis as CBAM~\cite{woo2018cbam} and concatenate them to produce an efficient feature descriptor, while~\cite{jiang2021bi} utilize CBAM mechanism on fused multi-scale to re-distribute the scale-wise weights.

\subsection{How to Use}
In Non-Transformer-based methods, the attention mechanism is typically used as a plug-in sublayer inserted into the convolutional block. The plug-in location can be divided into three categories: in the encoder, in the decoder, in the skip connection, and hybrid, based on where the attention layer occurs.

\begin{figure}[!h]
    \centering
    \centerline{\includegraphics[width=\linewidth]{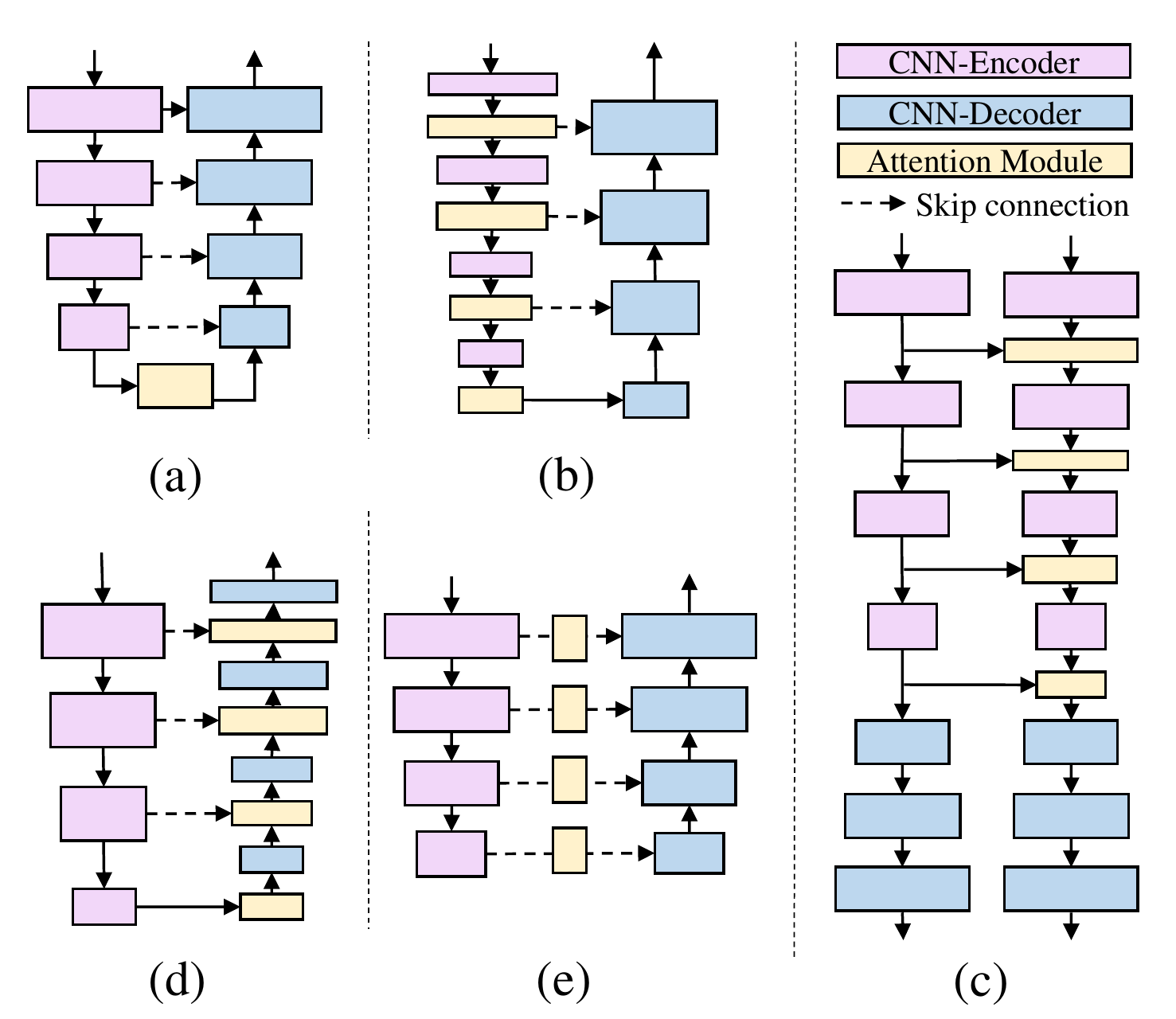}}
    \caption{Locations where the attention layer occurs in Non-Transformer-based methods, including (a) bottleneck (top of the encoder), (b)(c) each stage of the encoder in a network or between two networks, (d) decoder and (e) skip connection.}
    \label{attention-archit in segmentation}
\end{figure}

\subsubsection{In the Encoder}
The attention mechanism is often introduced into the encoder to enlarge the receptive field and extract richer encoded information.

\textbf{In the bottleneck.} Some researchers add the attention mechanism at the encoder's bottleneck, as shown in Fig~\ref{attention-archit in segmentation} of (a). 
With respect to the channel-based methods, Jiang \emph{et~al.} propose a scale-selection mechanism on concatenated multi-scale features, utilizing channel attention to calculate the correlation weight for each feature map with different scales to supplement the information lost in down-sampling.~\cite{yang2022automatic} propose an improvement on ECANet~(\cite{Wang_2020_CVPR}) by introducing a new branch that captures the distinct interactions between the current channel and its $k$ nearest channels after channel shuffling, thus enlarging the receptive fields. 
Differently, to highlight useful information and suppress redundant features, some adopt the spatial attention module(~\cite{cheng2022contour,guo2021sa,hu2019automatic,tang2019xlsor}), $e.g.$, CBAM~(\cite{woo2018cbam}) and the criss-cross attention~(\cite{huang2019ccnet}).~\cite{liu2019automatic,liu2021deep} create a pyramid-like feature structure~\cite{li2018pyramid} at the final output of the encoder, where~\cite{liu2019automatic} utilizes different convolution kernels and~\cite{liu2021deep} employs a pyramid local attention module to capture supporting information from compact and sparse ($i.e.$, different distance) neighboring contexts.
Moreover, since the convolution operations can only capture local information, self-attention is chosen to model global relations to solve this problem~(\cite{ding2019hierarchical,ding2020high,mou2019cs,fan2020ma}). In particular,~\cite{ding2019hierarchical} propose high-order attention and its improved version~\cite{ding2020high} to reduce computational costs, setting a threshold in the similarity matrix to reduce the noise of weak correlations. 
While~\cite{mou2019cs} apply channel-wise and spatial-wise self-attention in parallel to better aggregate features.~\cite{qu2022m3net} propose a dual-path network to utilize the complementation of arterial and venous phases visual information of CT and the attention module is inserted into the bottleneck of two networks to generate cross-phase reliable feature correlations as well as the channel dependencies.

%%%%%%%
\textbf{At each stage in a network.} To take advantage of multi-scale information, some works integrate the attention module into each stage in the encoder (Fig~\ref{attention-archit in segmentation} (b)). For example,~\cite{zhang2021caagp} propose a plug-and-play and lightweight attention module in each layer, which aggregates spatial information from $x$ and $y$ directions into channel attention through an adaptive global pooling.~\cite{singh2019fca} adopt a multi-scale input strategy~(\cite{van2017learning}), where the input images are resized into three different scales and the corresponding feature maps are aggregated as the encoder layer input. They also propose a factorized channel attention module with a 1-D kernel factorized convolution at each scale. 
Differently,~\cite{wang2019volumetric} construct a 2.5D network by combining the target slice with neighboring slices and apply both channel attention and spatial attention serially at each scale to obtain the corresponding response.

\textbf{At each stage between two networks.} Additionally, attention mechanisms are sometimes introduced to facilitate information transfer between two networks in the encoder, as illustrated in Fig~\ref{attention-archit in segmentation} (c). Those networks receive the same inputs.~\cite{min2019two} design a two-stream mutual attention network in semi-supervised segmentation to render the network robust in unclean data, whose input is the mixture of labeled data with manual labels and unlabeled data with pseudo labels. The features are exchanged in the mutual attention modules at each stage to discover incorrect labels and weaken their influence during parameter updating.~\cite{wang2019mixed} and~\cite{chen2019multi} propose multi-task hybrid supervised networks to improve segmentation performance. Specifically,~\cite{wang2019mixed} corrects the segmentation feature distribution using weak annotations from a detection network through attention modules, and~\cite{chen2019multi} leverages foreground and background information from the segmentation branch to build the auxiliary reconstruction task.

It is worth noting that~\cite{ahn2021multi} have proposed a novel spatiotemporal attention module in the encoder that can leverage the inter-frame consistency in echocardiography images. This module operates on a target slice and its neighboring frames and represents the first instance of spatiotemporal attention being applied to medical image segmentation.

% The attention mechanism applied in the encoder aims to integrate complementary information into encoded features, which operates as an element-wise multiplication or region selection at each scale or in the last scale.

\subsubsection{In the Decoder}
The decoder's primary task is to reconstruct the latent representation into an image that satisfies the specified criteria. Incorporating attention mechanisms in the decoder assists in generating dense feature maps by aggregating additional information, as demonstrated in (e) of Fig~\ref{attention-archit in segmentation}. These mechanisms can be classified based on their type of attention.

In the channel-wise applications,~\cite{ni2019rasnet} incorporate the attention module at each layer in the decoder to fuse multi-level features by utilizing the global context of high-level features as guidance information. Additionally, both~\cite{tomar2021ddanet} and~\cite{jia2019hd} utilize a dual-decoder architecture with one of the decoders acting as an auxiliary decoder to enhance segmentation. 
The attention mechanism is inserted in the main decoder to extract semantically discriminative intra-slice features in 3D prostate MRI segmentation~(\cite{jia2019hd}) or between two decoders to provide an attention map~(\cite{tomar2021ddanet}).

% It is quite common that the edge/shape information is of value in the spatial attention-based networks to reconstruct deep semantic features back to a high-resolution segmentation map. 
Spatial attention-based networks often benefit from incorporating edge/shape information to improve the reconstruction of deep semantic features into a high-resolution segmentation map. 
For instance,~\cite{karthik2022contour} use an attention module to capture contextual information from the contour feature maps in its spatial neighborhood.~\cite{wei2021shallow} use deep features to filter out background noise in shallow features and preserve edge information.~\cite{li2020joint} use the left atrial boundary as an attention mask on scar features to perform shape attention.~\cite{qin2020learning} propose an attention distillation technique to pass fine-grained details down to lower-resolution attention maps, improving their performance.

The channel-spatial attention modules are typically integrated into convolutional blocks in the decoder at each layer.  Especially,~\cite{fang2020spatial} found that combining the channel and spatial attention using element-wise addition achieved the best results in their experiments.~\cite{gu2020net} add another scale attention on the concatenated feature to achieve comprehensive attention based on the spatial-channel attention network.

Here, as experiments in~\cite{gu2020net} and~\cite{qin2020learning} have compared that the decoder-side attention block performs better than the encoder-side attention block. This may be because the encoders learn general feature, which is probably not strongly related to the target description. Thus applying an attention mechanism in the encoder may suppress some important information. However, this assumption needs to be further validated on a wider range of datasets with different modalities and under various scenarios to understand better the impact of attention mechanism placement on segmentation performance.
%This may be because suppressing some features early may limit the model's performance, as low-level features tend to capture local-scale information that is important for accurate segmentation \cite{gu2020net} and the low-level features in the encoder are more local-scale and not strongly related to target description~\cite{qin2020learning}.
% encoder 特征很重要，学通用的信息，但是和target关系可能不大，再encoder执行attention可能会把一些重要的信息抑制掉。所以aa和bb选择在decoder做attention。

\subsubsection{In the Skip Connection}
The skip connection plays a crucial role in restoring the full spatial resolution at the network's output, improving medical image segmentation performance.
The addition of the attention module (as shown in Fig~\ref{attention-archit in segmentation} (e)) can further enhance this process by mitigating the semantic gap between the encoder and decoder. Many examples in medical image segmentation demonstrate this.

The attention gate~(\cite{oktay2018attention}) is the most common form of attention mechanism in the skip connection, which is incorporated to selectively amplify informative features and suppress noise and irrelevant responses. 
Some researchers adopt the original spatial attention gate in their networks(\emph{e.g.,}~\cite{zuo2021r2au,li2020anu,kearney2019attention,wu2019u,li2020attention}), while others keep the form with different attention computations. 
For instance,~\cite{khanh2020enhancing} improve it with channel-spatial attention instead. 
Inspired by the focal loss,~\cite{yeung2021focus} add a focal parameter to reduce the contributions of easy examples for harder examples learning.~\cite{li2019connection} also incorporate the loss function into their network design by designing a topology-aware loss. This loss function aims to model the structure properties by encouraging the probability of connectivity in the neighboring region to be high.

The attention module can also be inserted into the skip connection at each stage in a different way than the attention gate. 
At first, well-known attention methods are applied to the skip connection(~\cite{yang2021improved,hu2021sa,sinha2020multi,cheng2020fully,zhou2021automatic,guo2020residual,guo2021channel}), $e.g.$, SE~(\cite{hu2018squeeze}), CBAM~(\cite{woo2018cbam}), Efficient channel attention(ECA, \cite{Wang_2020_CVPR}), Concurrent Spatial and Channel Squeeze \& Excitation (scSE, \cite{roy2018concurrent}), self-attention~(\cite{vaswani2017attention}). 
Besides, the attention module is improved for various purposes. To respond to the different directions of the features for better information conservation,~\cite{tong2021sat} design a side attention block through different shapes of convolution kernels. 
To bring in multi-scale property for its richer semantic information,~\cite{xia2022mc} input the multi-scale encoded feature to the skip connection and then apply residual attention to highlight salient areas for each stage. Differently,~\cite{wang2021hybrid} and~\cite{lyu2020attention} design multi-scale attention directly to solve the problem.~\cite{wang2021hybrid} applies channel attention at each layer and only adds additional hybrid dilated attention layers with different dilation rates at the bottleneck.~\cite{lyu2020attention} inserts the multi-scale attention module at each layer of their network, using a multi-branch architecture with different numbers of convolutional operations to capture semantic information without reducing dimensions.
Additionally, some researchers consider that features from different levels should be handled differently due to their characteristics.~\cite{tong2019rianet} perform spatial attention on low-level features and channel attention on high-level features.~\cite{fan2020pranet} and~\cite{lou2021caranet} only apply the attention mechanism to high-level features to reduce computational costs. Reverse attention is designed in the former to establish the relationship between areas and boundary cues, and axial reverse attention is designed in the latter to analyze localization information.

\subsubsection{Hybrid}
To combine the advantages of the methods mentioned above and apply various attention to the network, some researchers integrate attention into multiple locations in the architecture design. 

Among them, the most common form is inserting the attention module into both the encoder and decoder in each layer with different attention modules.
% That is, the attention module is inserted into the convolution block for the encoder and decoder.
The attention modules, like CBAM~(\cite{guo2020residual}), SE block~(\cite{yin2022sd}) and its varieties~(\cite{xie2021duda,li2021bseresu,wang2022attention}), are usually adopt. 
Moreover, since SE block and its varieties recalibrate spatial information for the same single weight,~\cite{lu2022fine} propose double group attention modules to extract multigroup weighs of the feature maps to strengthen the spatial information. 
Furthermore,~\cite{xu2021exploiting} consider introducing long-term dependency to exploit context prior. 
Some works also take the functional differences between the encoder and decoder and apply distinct attention mechanisms. 
The encoder is typically designed to extract features from images while preserving as much detail as possible, while the decoder is responsible for recovering enough information from the features to enable accurate segmentation. As a result, it is common practice to apply channel attention to the encoder to enhance the representation of target features, while using spatial attention in the decoder to emphasize the position of useful areas when fusing low-level and high-level features. This approach has been demonstrated in various studies, such as~\cite{li2021ta} and~\cite{guo2021dual}.
The attention module can also be applied between two frameworks for information transactions in the encoder and decoder. 
Some works adopt the same inputs but different networks design. Due to the task relationship brought inherently by the data,~\cite{xu2019deep} utilize the properties of the progressive inclusion of subregions to guide the current task with the outputs of previous tasks as salient regions on the BraTS dataset. In~\cite{xu2020asymmetrical}, the connection between the prostate bed and the bladder and rectum segmentation task is considered, and the feature from the auxiliary network is transferred to the target prostate bed segmentation network through the attention module.
Besides, the attention mechanism is also integrated into the skip connection based on the encoder-decoder attention architecture~(\cite{yao2022novel}). 

However, there is currently no experiment that demonstrates the individual effects of adding attention at different locations on performance gain.

\subsection{Where to use (Applications)}

After discussing the basic types and the embedded locations of attention mechanisms used in medical image segmentation, we now introduce the specific application tasks. 
These tasks mainly include brain, breast, cardiac, kidney, liver, lung, polyp, prostate, eye, skin lesion, and surgical instrument segmentation. Some universal attention-based medical segmentation models may not be designed for specific tasks but for multiple organs/lesions segmentations on different modalities datasets as shown in Table~\ref{universal} 
In addition, we also give a series of tables to illustrate applications on specific organs with attention types (What to use), locations (How to use), and their performance on the common dataset. 
% We conclude the main tasks and the mainly used attention types, and introduce some organ-specific attention design if possible. Besides, we also list some tables to illustrate applications on specific organ with attention types, locations and their performance on the common dataset.

%\textbf{Universal.}
%The attention-based methods designed for universal segmentation are more challenging than organ-specific segmentation for its uncertainty and unknown tasks. Thus we can conclude that applying spatial attention in the skip connections are the most popular type to obtain more edge/texture information and the experiments are demonstrated on various datasets as shown in Table~\ref{universal}.
%\xie{Owing to the uncertainty and unknown tasks, attention-based methods designed for universal segmentation are more challenging than organ-specific segmentation. In Table~\ref{universal}, we list advanced attention-based universal medical image segmentation methods. It shows that the commonly used attention types are spatial attention and spatial-channel attention. The popular location is intermediate.}

% Thus  are the most popular attention type to obtain more edge/texture information and the experiments are demonstrated on various datasets as shown in Table~\ref{universal}.
% 

\renewcommand\arraystretch{1}
\begin{table*}[!t]
\caption{An overview of non-Transformer methods for universal segmentation. `How/What' means `How/What to use'.}
\label{universal}
%\vspace{-0.1cm} 
\resizebox{\linewidth}{!}{
\begin{tabular}{@{}lllllll@{}}
\toprule
Author & How & What & Datasets     & Metric                                            & \begin{tabular}[c]{@{}l@{}}Data split \\ (Train:Val:Test)\end{tabular} & Modality (Type) \\ \midrule
\multirow{2}{*}{\cite{wang2019mixed}}     & \multirow{2}{*}{Skip.} & \multirow{2}{*}{Channel} & 1.Lung nodule dataset (NM)                                                               & Dice:0.8490                                                                        & 160:80:80                        & \multirow{2}{*}{CT (2D)} \\ \cdashline{4-6}[0.8pt/5pt]
               &                         &                          & 2.Inner ear dataset (NM)                                                                 & Dice:0.8873                                                                        & 66:40:40                         &                          \\ \cmidrule(l){4-7}
\multirow{2}{*}{\cite{kaul2019focusnet}} & \multirow{2}{*}{Skip.}    & \multirow{2}{*}{Spatial} & 1.ISBI 2017 ~(\cite{codella2018skin})                                                                         & \begin{tabular}[c]{@{}l@{}}ACC:0.9214 JI:0.7562 Dice:0.8315\\SE:0.7673\end{tabular} & 2000:150:600                          & Dermoscopy (2D)            \\ \cdashline{4-7}[0.8pt/5pt]
                 &                       &                          & 2.\href{https://www.kaggle.com/datasets/kmader/finding-lungs-in-ct-data}{Finding and Measuring Lungs in CT Data} & ACC:0.9932 JI:0.9965 SE:0.9757                                                      & 80\%:20\%:NM                               & CT (2D)                    \\ \cmidrule(l){4-7} 
\multirow{5}{*}{\cite{zhang2019net}} & \multirow{5}{*}{Skip.}   & \multirow{5}{*}{Spatial} & 1.REFUGE ~(\cite{orlando2020refuge})                                                                           & Dice:0.8912(OC) 0.9529(OD)                                                              & 400:400:NaN                      & Retinal fundus (2D)    \\ \cdashline{4-7}[0.8pt/5pt]
            &                            &                          & 2.Drishti-GS ~(\cite{sivaswamy2014drishti})                                                                             & Dice:0.9314(OC) 0.9752(OD)                                                              & 50:51:NaN                        & Retinal fundus (2D)    \\ \cdashline{4-7}[0.8pt/5pt] 
            &                            &                          & 3.DRIVE ~(\cite{staal2004ridge})                                                                           & ACC:0.9560 mIoU:0.7744                                                             & 20:NaN:20                        & Retinal fundus (2D)    \\ \cdashline{4-7}[0.8pt/5pt] 
            &                            &                          & 4.MC ~(\cite{jaeger2014two})                                     & ACC:0.9865 mIoU:0.9420                                                             & 80:NaN:58                        & X-Ray (2D)             \\ \cdashline{4-7}[0.8pt/5pt]
             &                           &                          & 5.\href{https://www.kaggle.com/datasets/kmader/finding-lungs-in-ct-data}{Finding and Measuring Lungs in CT Data}                                                                   & ACC:0.9868 mIoU:0.9623                                                             & 214:NaN:53                       & CT (2D)                \\ \cmidrule(l){4-7} 
\multirow{4}{*}{\cite{lin2020refineu}}  & \multirow{4}{*}{Skip.}    & \multirow{4}{*}{Spatial} & 1.ETIS~(\cite{silva2014toward})                                                                              & Dice:0.9602 JI:0.8133 ACC:0.9936                                                   & 156:NaN:40                       & Endoscopy (2D)         \\ \cdashline{4-7}[0.8pt/5pt]
                &                        &                          & 2.CVC-ColonDB~(\cite{bernal2012towards})                                                                           & Dice:0.9386 JI:0.8234 ACC:0.9823                                                   & 304:NaN:76                       & Endoscopy (2D)         \\ \cdashline{4-7}[0.8pt/5pt] 
           &                             &                          & 3.ISBI 2016~(\cite{gutman2016skin})                                                                         & Dice:0.9172 JI:0.8508 ACC:0.9823                                                   & 900:NaN:379                      & Dermoscopy (2D)        \\ \cdashline{4-7}[0.8pt/5pt] 
            &                            &                          & 4.ISBI 2017~(\cite{codella2018skin})                                                                         & Dice:0.8775 JI:0.7538 ACC:0.9321                                                   & 2000:NaN:600                     & Dermoscopy (2D)        \\ \cmidrule(l){4-7}
\multirow{3}{*}{\cite{ding2020high} } & \multirow{3}{*}{Skip.}    & \multirow{3}{*}{Spatial} & 1.REFUGE~(\cite{orlando2020refuge})                                                                            & mDice:0.9302                                                                        & 400:400:NaN                      & Retinal fundus (2D)    \\ \cdashline{4-7}[0.8pt/5pt]
            &                            &                          & 2.\href{https://www.kaggle.com/datasets/kmader/finding-lungs-in-ct-data}{Finding and Measuring Lungs in CT Data}                                                                              & ACC:0.9945 Sen:0.9879 mIoU:0.9849                                                             & 214:NaN:53                       & CT (2D)                \\ \cdashline{4-7}[0.8pt/5pt] 
        &                                &                          & 3.DRIVE ~(\cite{staal2004ridge})                                                                             & \begin{tabular}[c]{@{}l@{}}ACC:0.9712 F1:0.8300 SE:0.8297 \\ SP:0.9843 mIoU:0.7094\end{tabular}                                                               & 20:NaN:20                   & Retinal fundus (2D)    \\ \cmidrule(l){4-7} 
\multirow{3}{*}{\cite{zuo2021r2au} } & \multirow{3}{*}{Skip.}     & \multirow{3}{*}{Spatial} & 1.CVC- 2018 ~(\cite{tschandl2018ham10000})                                                                              & ACC:0.9277 F1:0.8660 AUC:0.8957                                                               & 1815:259:520                     & Dermoscopy (2D)        \\ \cdashline{4-7}[0.8pt/5pt]
                &                        &                          & 2.DRIVE ~(\cite{staal2004ridge})                                                                             & ACC:0.9555 F1:0.8213 AUC:0.9790                                                               & 20:NaN:20                 & Retinal fundus (2D)    \\ \cdashline{4-7}[0.8pt/5pt]
           &                             &                          & 3.\href{https://www.kaggle.com/datasets/kmader/finding-lungs-in-ct-data}{Finding and Measuring Lungs in CT Data}                                                                             & ACC:0.9950 F1:0.9868 AUC:0.9921                                                               & 134:54:79                        & CT (2D)                \\ \cmidrule(l){4-7}
\multirow{2}{*}{\cite{an2021medical}}  & \multirow{2}{*}{Skip.}     & \multirow{2}{*}{Spatial} & 1.IBSR ~(\cite{IBSR})                                                                             & Jaccard similarity:0.9725                                                          & NM                               & MRI (2D)               \\ \cdashline{4-7}[0.8pt/5pt]
           &                             &                          & 2.JSRT ~(\cite{rikitake2019use})                                                                              & Jaccard similarity:0.9836                                                          & NM                       & CT (2D)                \\ \cmidrule(l){4-7}
\multirow{4}{*}{\cite{xia2022mc}}  & \multirow{4}{*}{Skip.}    & \multirow{4}{*}{Spatial} & 1.\href{https://www.kaggle.com/datasets/kmader/finding-lungs-in-ct-data}{Finding and Measuring Lungs in CT Data}                                                               & ACC:0.9960 AUC:0.9947 Dice:0.9897                                                    & \multirow{4}{*}{80\%:NaN:20\%} & \multirow{4}{*}{CT (2D)} \\ \cdashline{4-5}[0.8pt/5pt]
           &                             &                          & 2.Bladder dataset (Private dataset)                                                                           & ACC:0.9946 AUC:0.9679 Dice:0.9679                                                    &                                    &                          \\ \cdashline{4-5}[0.8pt/5pt]
           &                             &                          & 3.\href{https://www.tipdm.org/bdrace/tzjingsai/20181226/1544.html}{Tipdm Cup Challenge}                                                               & ACC:0.9981 AUC:0.9367 Dice:0.9977                                                    &                                    &                          \\ \cdashline{4-5}[0.8pt/5pt]
           &                             &                          & 4.KiTs ~(\cite{heller2019kits19})                                                                             & ACC:0.9924 AUC:0.9356 Dice:0.9729                                                    &                                    &                          \\ \cmidrule(l){4-7}
\multirow{4}{*}{\cite{cheng2022contour}} & \multirow{4}{*}{Skip.}   & \multirow{4}{*}{Spatial} & 1.\begin{tabular}[c]{@{}l@{}}COVID-19 CT segmentation dataset ~(\cite{medicalsegmentation.com})\end{tabular}                                                                          & F1:0.7516 AUC:0.8382                                                               & 45:5:50                          & CT (2D)                \\ \cdashline{4-7}[0.8pt/5pt] 
            &                            &                          & 2.CVC-ClinicDB ~(\cite{bernal2015wm})                                                                      & F1:0.7998 AUC:0.8804                                                               & 414:85:113                       & Endoscopy (2D)         \\ \cdashline{4-7}[0.8pt/5pt] 
            &                            &                          & 3.CVC- 2018 ~(\cite{tschandl2018ham10000})                                                                          & F1:0.8627 Dice:0.9041                                                              & 1815:259:520                     & Dermoscopy (2D)        \\ \cdashline{4-7}[0.8pt/5pt] 
            &                            &                          & 4.\begin{tabular}[c]{@{}l@{}}Lung nodule competition 2017 ~(\cite{liao2019evaluate})\end{tabular}                                                      & F1:0.9868 AUC:0.9929                                                               & 571:143:307                      & CT (2D)                \\ \cmidrule(l){4-7}
\multirow{2}{*}{\cite{min2019two} } &
  \multirow{2}{*}{Skip.} &
  \multirow{2}{*}{\begin{tabular}[c]{@{}l@{}}Spatial and \\ Channel\end{tabular}} &
  1.HVMSR 2016 ~(\cite{pace2015interactive}) &
  Overall score:-0.024 &
  10:NaN:10 &
  \multirow{2}{*}{MRI (3D)} \\ \cdashline{4-6}[0.8pt/5pt]
 &  &
   &
  2.BRATS-2015 ~(\cite{kistler2013virtual}) &
  Dice:0.792 &
  244:NaN:30 &
   \\ \cmidrule(l){4-7}
\multirow{3}{*}{\cite{sinha2020multi}} &
  \multirow{3}{*}{Skip.} &
  \multirow{3}{*}{\begin{tabular}[c]{@{}l@{}}Spatial and \\ Channel\end{tabular}} &
  1.CHAOS ~(\cite{CHAOS2021}) &
  Dice:0.8675 MSD:0.66 &
  13:2:5 &
  \multirow{3}{*}{MRI (3D)} \\ \cdashline{4-6}[0.8pt/5pt]
 &  &
   &
  2.HSVMR 2016 ~(\cite{pace2015interactive}) &
  Dice:0.8320 MSD:1.19 &
  60\%:20\%:20\% &
   \\ \cdashline{4-6}[0.8pt/5pt]
 &  &
   &
  3.\begin{tabular}[c]{@{}l@{}}Brain segmentation dataset of MSD ~(\cite{antonelli2022medical})\end{tabular} &
  Dice:0.8037 MSD:0.90 &
  388:48:48 &
   \\ \cmidrule(l){4-7}
\multirow{3}{*}{\cite{khanh2020enhancing} } &
  \multirow{3}{*}{Skip.} &
  \multirow{3}{*}{\begin{tabular}[c]{@{}l@{}}Spatial and \\ Channel\end{tabular}} &
  1.CVC-ClinicDB ~(\cite{bernal2015wm}) &
  Dice:0.7331 &
  \multirow{3}{*}{80\%:20\%:NaN} &
  Endoscopy (2D) \\ \cdashline{4-5}[0.8pt/5pt] \cdashline{7-7}[0.8pt/5pt] 
 &  &
   &
  2.VIP-VUP18 ~(\cite{VIP-VUP18}) &
  Dice:0.5626 &
   &
  CT (2D) \\ \cdashline{4-5}[0.8pt/5pt] \cdashline{7-7}[0.8pt/5pt] 
 &  &
   &
  3.TCGA ~(\cite{buda2019association}) &
  Dice:0.8583 &
   &
  MRI (2D) \\ \cmidrule(l){4-7}
\multirow{3}{*}{\cite{cheng2020fully}} &
  \multirow{3}{*}{Skip.} &
  \multirow{3}{*}{\begin{tabular}[c]{@{}l@{}}Spatial and \\ Channel\end{tabular}} &
  1.Chest X-ray collection ~(\cite{demner2016preparing})&
  P:0.9860 IoU:0.9669 Dice:0.9832 &
  \multirow{3}{*}{80\%:10\%:10\%} &
  X-ray (2D) \\ \cdashline{4-5}[0.8pt/5pt] \cdashline{7-7}[0.8pt/5pt] 
 &  &
   &
  \begin{tabular}[c]{@{}l@{}}2.Kaggle 2018 data science bowl ~(\cite{hamilton2018kaggle})\end{tabular} &
  P:0.9022 IoU:0.8209 Dice:0.8989 &
   &
  Retinal fundus (2D) \\ \cdashline{4-5}[0.8pt/5pt] \cdashline{7-7}[0.8pt/5pt] 
 &  &
   &
  3.Herlev ~(\cite{zhao2020pgu,zhang2017deeppap}) &
  P:0.9413 IoU:0.8862 Dice:0.9321 &
   &
  Mircoscopy (2D) \\ \cmidrule(l){4-7}
\multirow{2}{*}{\cite{fang2020spatial} } &
  \multirow{2}{*}{Dec} &
  \multirow{2}{*}{\begin{tabular}[c]{@{}l@{}}Spatial and \\ Channel\end{tabular}} &
  1.\href{https://www.kaggle.com/datasets/avc0706/luna16}{LUNA} &
  F1:0.9841 AUC:0.9897 &
  70\%:NaN:30\% &
  CT (2D) \\ \cdashline{4-7}[0.8pt/5pt] 
 &  &
   &
  2.ISIC ~(\cite{tschandl2018ham10000}) &
  F1:0.8700 AUC:0.9310 &
  1815:259:520 &
  Dermoscopy (2D) \\ \cmidrule(l){4-7}
\multirow{2}{*}{\cite{gu2020net}} &
  \multirow{2}{*}{Dec} &
  \multirow{2}{*}{\begin{tabular}[c]{@{}l@{}}Spatial and \\ Channel\end{tabular}} &
  1.ISIC 2018 ~(\cite{tschandl2018ham10000}) &
  Dice:0.9208 &
  1816:260:518 &
  Dermoscopy (2D) \\ \cdashline{4-7}[0.8pt/5pt] 
 &  &
   &
  \begin{tabular}[c]{@{}l@{}}2.Placenta and Fetal Brain Segmentation (Private dataset)\end{tabular} &
  \begin{tabular}[c]{@{}l@{}}Dice:0.8708 (placenta), 0.9588 (brain)\end{tabular} &
  1050:150:300 &
  MRI (2D) \\ \cmidrule(l){4-7}
\multirow{3}{*}{\cite{gao2021multiscale}} &
  \multirow{3}{*}{Skip.} &
  \multirow{3}{*}{\begin{tabular}[c]{@{}l@{}}Spatial and \\ Channel\end{tabular}} &
  1.BUS ~(\cite{yap2017automated}) &
  Dice:0.8539 IoU:0.7706 P:0.8831 &
  NM &
  Ultrasound (2D) \\ \cdashline{4-7}[0.8pt/5pt] 
 &  &
   &
  2.CVC-ClinicDB ~(\cite{bernal2015wm}) &
  Dice:0.8535 IoU:0.7861 P:0.8950 &
  NM &
  Endoscopy (2D) \\ \cdashline{4-7}[0.8pt/5pt] 
 &  &
   &
  3.ISBI-2014 ~(\cite{lu2016evaluation, lu2015improved}) &
  Dice:0.9082 IoU:0.8331 P:0.8822 &
  45:NaN:90 &
  Microscopy (2D) \\ \cmidrule(l){4-7}
\multirow{2}{*}{\cite{xu2021exploiting} } &
  \multirow{2}{*}{Enc \& Dec} &
  \multirow{2}{*}{\begin{tabular}[c]{@{}l@{}}Spatial and \\ Channel\end{tabular}} &
  1.Fetal A4C (Private dataset)&
  Dice:0.849 HD:3.421 &
  \multirow{2}{*}{70\%:NaN:30\%} &
  \multirow{2}{*}{Ultrasound (2D)} \\ \cdashline{4-5}[0.8pt/5pt]
 &  &
   &
  2.Fetal head ~(\cite{van2018automated}) &
  Dice:0.971 HD:3.234 &
   &
   \\ \cmidrule(l){4-7}
\multirow{2}{*}{\cite{yao2022novel}} &
  \multirow{2}{*}{\begin{tabular}[c]{@{}l@{}}Enc, Dec \\ \& Skip.\end{tabular}} &
  \multirow{2}{*}{\begin{tabular}[c]{@{}l@{}}Spatial and \\ Channel\end{tabular}} &
  \begin{tabular}[c]{@{}l@{}}1.Vestibular schwannoma segmentation dataset~(\cite{shapey2021segmentation})\end{tabular} &
  mDice:0.8233 &
  \multirow{2}{*}{NM} &
  MRI (3D) \\ \cdashline{4-5}[0.8pt/5pt] \cdashline{7-7}[0.8pt/5pt] 
 &  &
   &
  2.BCV ~(\cite{landman2015miccai}) + CHAOS ~(\cite{kavur2021chaos}) &
  mDice:0.8704 (MRI-CT), 0.9125 (CT-MRI) &
   &
  CT, MRI (3D) \\
\bottomrule
\end{tabular}
}
\end{table*}

\textbf{Brain.}
The brain segmentation applications focus on the brain tissue ~(\cite{sun20193d}) and tumor segmentation task ~(\cite{noori2019attention,yuan2019unified,islam2019brain,chen2019multi,zhou2018learning,zhou2020one,akil2020fully,zhou2020multi,xu2019deep,zhang2020attention,maji2022attention,fang2022nonlocal}), here we list the applications on glioma segmentation as it is the mainstream due to the muli-modal brain tumor segmentation benchmark (BraTS, \cite{menze2014multimodal}) (shown in Table~\ref{brain}). Especially, most of the attention designs are based on SE blocks~(\cite{hu2018squeeze}) and integrated into the decoder. 
Some works consider that unrelated information such as the contour as well as internal structures of the brain may be preserved in the encoder and insert the attention layer into the encoder to emphasize tumor-related features~(\cite{yuan2019unified,noori2019attention}). 
Others take into account the potential subregion correlations between the whole tumor (WT), enhanced tumor (ET), and tumor core (TC) and propose a cascaded attention mechanism in the decoder~(\cite{xu2019deep,zhou2020one}). 
In a semi-supervised approach proposed by~\cite{chen2019multi}, labels generated using an attention mechanism are used to help the reconstruction task separate foreground and background, thus aiding in the segmentation of brain tumors and white matter hyperintensities due to the shared encoder.

\begin{table*}[!h]
%\scriptsize
\renewcommand\arraystretch{1.25} %1.25表示1.25倍行高
\caption{An overview of non-Transformer methods for brain segmentation, in which WT, TC, ET, WM, GM, and CSF is the whole tumor, tumor core, enhanced tumor, white matter, gray matter, and cerebrospinal fluid seperately.}
\label{brain}
%\vspace{-0.5cm}
%\begin{center}
\resizebox{\linewidth}{!}{
\begin{tabular}{@{}lllllllll@{}}
\toprule
Author  & How & What & Datasets & \multicolumn{3}{c}{Metric}  & Data split(Train:Val:Test) & Modality (Type) \\  \midrule

\multirow{2}{*}{\cite{noori2019attention}} &
  \multirow{2}{*}{Dec} &
  \multirow{2}{*}{Channel} &
  1.BRATS 2017 ~(\cite{menze2014multimodal, bakas2017advancing})&
  Dice:0.791 (ET) & 0.885 (WT) & 0.783 (TC) &
  \multirow{2}{*}{NM} &
  \multirow{2}{*}{MRI (2D)} \\ \cdashline{4-7}[0.8pt/5pt]
  &     &           & 2.BRATS 2018 ~(\cite{menze2014multimodal, bakas2017advancing}) & Dice:0.813 (ET) & 0.895 (WT) & 0.823 (TC) &                            &                 \\ \cmidrule(l){4-9} 
\multirow{2}{*}{\cite{zhou2018learning}} &
  \multirow{2}{*}{Enc \& Dec} &
  \multirow{2}{*}{Channel} &
  \multirow{2}{*}{BRATS2018~(\cite{menze2014multimodal, bakas2017advancing})} &
  Dice:0.7775 (ET) & 0.8842 (WT) & 0.7960 (TC) &
  \multirow{2}{*}{NM} &
  \multirow{2}{*}{MRI (3D)} \\ \cdashline{5-7}[0.8pt/5pt]
  &     &           &  & HD95:2.9366 (ET) & 5.4681 (WT) & 6.8773 (TC)           &                            &                 \\ \cmidrule(l){4-9} 
\multirow{2}{*}{\cite{sun20193d}} &
  \multirow{2}{*}{Enc \& Dec} &
  \multirow{2}{*}{Spatial} &
  1.MRBrainS13 ~(\cite{mendrik2015mrbrains}) &
  Dice:0.8656 (GM) & 0.8886 (WM) & 0.8553 (CSF) &
  80\%:NaN:20\% &
  \multirow{2}{*}{MRI (3D)} \\ \cdashline{4-8}[0.8pt/5pt]
  &     &           & 2.MALC12 ~(\cite{landman2012miccai})    & Dice:0.8482    &  &                                     & 20:NaN:10                &                 \\ \cmidrule(l){4-9} 
\cite{yuan2019unified} &
  Dec &
  Spatial &
  Brain Tumors Task of MSD ~(\cite{antonelli2022medical}) &
  Dice:0.5361 (T1Gd) & 0.8155 (FLARI) & 0.7654 (T2)&
  50\%:NaN:50\% &
  MRI (2D) \\ \cmidrule(l){4-9} 
\multirow{2}{*}{\cite{chen2019multi}} &
  \multirow{2}{*}{Skip.} &
  \multirow{2}{*}{Spatial} &
  1.BRATS 2018 ~(\cite{menze2014multimodal, bakas2017advancing}) &
  Dice:0.7702  &  & &
  120:50:50  &
  \multirow{2}{*}{MRI (2D)} \\ \cdashline{4-8}[0.8pt/5pt]
  &     &           & 2.WMH17 ~(\cite{kuijf2019standardized})     & Dice:0.7204      &  &                                   & 30:10:20                 &                 \\ \cmidrule(l){4-9} 
\cite{islam2019brain} &
  Dec &
  \begin{tabular}[c]{@{}l@{}}Spatial and \\ Channel\end{tabular} &
  BRATS 2019 ~(\cite{menze2014multimodal, bakas2017advancing, bakas2018identifying}) &
  Dice:0.7780 (ET) & 0.8689 (WT) & 0.7771 (TC) & 
  NM &
  MRI (3D) \\ \cmidrule(l){4-9} 
\cite{xu2019deep}  & 
  Dec & 
  Spatial & 
  BRATS 2018 ~(\cite{menze2014multimodal, bakas2017advancing}) & 
  Dice: 0.8171 (ET) & 0.9118 (WT) & 0.8619 (TC) & 
  NM & 
  MRI (3D) \\ \cmidrule(l){4-9} 
\multirow{3}{*}{\cite{zhou2020one} } &
  \multirow{3}{*}{Dec} &
  \multirow{3}{*}{Channel} &
  1.BRATS 2018 ~(\cite{menze2014multimodal, bakas2017advancing}) &
  Dice:0.8111 (ET) & 0.9078 (WT) & 0.8575 (TC) &
  \multirow{3}{*}{NM} &
  \multirow{3}{*}{MRI (3D)} \\ \cdashline{4-7}[0.8pt/5pt]
  &     &           & 2.BRATS 2017~(\cite{menze2014multimodal, bakas2017advancing}) & Dice:0.7852 (ET) & 0.9071 (WT) & 0.8422 (TC) &                            &                 \\ \cdashline{4-7}[0.8pt/5pt]
  &     &           & 3.BRATS 2015~(\cite{menze2014multimodal}) & Dice:0.65 (ET) & 0.87 (WT) & 0.75 (TC) &                            &                 \\ \cmidrule(l){4-9} 
\multirow{2}{*}{\cite{maji2022attention}} &
  \multirow{2}{*}{Skip.} &
  \multirow{2}{*}{Spatial} &
  \multirow{2}{*}{BRATS2019~(\cite{menze2014multimodal, bakas2017advancing, bakas2018identifying})} &
  Dice:0.801 (ET) & 0.911 (WT) & 0.876 (TC) & 
  \multirow{2}{*}{4700:775:1000} &
  \multirow{2}{*}{MRI (2D)} \\ \cdashline{5-7}[0.8pt/5pt]
    &     &           &   & mIoU:0.668 (ET) & 0.838 (WT) & 0.781 (TC) &  & \\
\bottomrule
\end{tabular}
}
\end{table*}

\textbf{Breast.}
The segmentation tasks on breast focus on breast anatomy~(\cite{lei2020self}), breast cancer~(\cite{lee2020channel,zhuang2019rdau,vakanski2020attention,punn2022rca}) and breast mass~(\cite{li2019attention,sun2020aunet}) as shown in Table~\ref{breast}. Through this literature, we can conclude that most attention methods used in this domain are directly transplanted, $e.g.$, SE blocks~(\cite{hu2018squeeze}), attention gate~(\cite{oktay2018attention}). Some works also attempt to introduce domain knowledge. For instance,~\cite{vakanski2020attention} obtain the saliency map~(\cite{xu2021tumor}) through prior knowledge constraints (the information of the degree of connectedness and confidence of connectedness between the image regions) and utilize precomputed saliency maps in the attention module that point out to target spatial regions. 

\begin{table*}[!h]
%\scriptsize
\renewcommand\arraystretch{1.25} %1.25表示1.25倍行高
\caption{An overview of non-Transformer methods for breast segmentation.}
\label{breast}
%\vspace{-0.5cm}
%\begin{center}
\resizebox{\linewidth}{!}{
\begin{tabular}{@{}lllllll@{}}
\toprule
Author & How & What & Datasets     & Metric                                            & \begin{tabular}[c]{@{}l@{}}Data split \\ (Train:Val:Test)\end{tabular} & Modality (Type) \\ \midrule
\multirow{3}{*}{\cite{zhuang2019rdau}} &
  \multirow{3}{*}{Skip.} &
  \multirow{3}{*}{Spatial} &
  \begin{tabular}[c]{@{}l@{}}1.\href{ultrasoundcases.info}{ultrasoundcases}\end{tabular} &
  \multirow{3}{*}{\begin{tabular}[c]{@{}l@{}}ACC:0.9791 Dice:0.8469 F1:0.8478 \\ IoU:0.8067 AUC:0.9227\end{tabular}} &
  NaN:NaN:100\% &
  \multirow{3}{*}{Ultrasound (2D)} \\ \cdashline{4-4}[0.8pt/5pt] \cdashline{6-6}[0.8pt/5pt]
 &  &
   &
  \begin{tabular}[c]{@{}l@{}}2.Private dataset\end{tabular} &
   &
  NaN:NaN:100\% &
   \\ \cdashline{4-4}[0.8pt/5pt] \cdashline{6-6}[0.8pt/5pt]
 &  &
   &
  \begin{tabular}[c]{@{}l@{}}3.Breast ultrasound dataset ~(\cite{yap2017automated})\end{tabular} &
   &
  730:127:NaN &
   \\ \cmidrule(l){4-7} 
\cite{li2019attention} &
  Skip. &
  Spatial &
  DDSM ~(\cite{pubdigital}) &
  F1:0.8224 SE:0.7789 &
  66\%:17\%:17\% &
  manmography (2D) \\ \cmidrule(l){4-7} 
\cite{vakanski2020attention} &
  Enc &
  Spatial &
  BUSIS ~(\cite{xian2018benchmark}) &
  Dice:0.901 JI:0.832 ACC:0.979 AUC:0.955 &
  80\%:20\%:NaN &
  Ultrasound (2D) \\ \cmidrule(l){4-7} 
\cite{lee2020channel} &
  Enc &
  Channel &
  \begin{tabular}[c]{@{}l@{}}Breast ultrasound dataset ~(\cite{yap2017automated})\end{tabular} &
  ACC:0.97794 F1:0.7658 IoU:0.6226 &
  146/147:NaN:16/17 &
  Ultrasound (2D) \\ \cmidrule(l){4-7} 
\multirow{2}{*}{\cite{lei2020self}} &
  \multirow{2}{*}{Enc} &
  \multirow{2}{*}{\begin{tabular}[c]{@{}l@{}}Spatial and \\ Channel\end{tabular}} &
  1.Private dataset &
  Dice:0.866 P:0.954 ACC:0.922 &
  \multirow{2}{*}{NM} &
  \multirow{2}{*}{Ultrasound (2D)} \\ \cdashline{4-5}[0.8pt/5pt]
 &  &
   &
  2.Private dataset &
  Dice:0.911 P:0.912 ACC:0.963 &
   &
   \\ \cmidrule(l){4-7} 
\multirow{2}{*}{\cite{sun2020aunet} } &
  \multirow{2}{*}{Dec} &
  \multirow{2}{*}{Channel} &
  1.CBIS-DDSM ~(\cite{pubdigital, lee2017curated}) &
  Dice:0.818 SE:0.849 &
  690:168:NaN &
  \multirow{2}{*}{Mammography (2D)} \\ \cdashline{4-6}[0.8pt/5pt]
  &  &
   &
  2.Inbreast ~(\cite{moreira2012inbreast}) &
  Dice:0.791 SE:0.808 &
  80\%:20\%:NaN &
   \\ \cmidrule(l){4-7} 
\multirow{2}{*}{\cite{punn2022rca}} &
  \multirow{2}{*}{Skip.} &
  \multirow{2}{*}{Spatial} &
  1.BUSIS ~(\cite{xian2018benchmark}) &
  ACC:0.990 Dice:0.937 mIoU:0.910 &
  \multirow{2}{*}{70\% : NaN : 30\%} &
  \multirow{2}{*}{Ultrasound (2D)} \\ \cdashline{4-5}[0.8pt/5pt]
  &  &
   &
  2.BUSI ~(\cite{al2020dataset}) &
  ACC:0.970 Dice:0.914 mIoU:0.899 &
   &
   \\
\bottomrule
\end{tabular}
}
\end{table*}

\textbf{Cardiac.}
The segmentation tasks surrounding the heart mainly include left ventricle~(\cite{ge2019k,ahn2021multi,lu2022fine}), left atrial~(\cite{li2020joint}), and cardiac anatomical structure segmentation~(\cite{tong2019rianet,guo2021dual,liu2021deep,liu2021automated,wang2022awsnet,ding2020cab}) and we list the cardiac anatomical structure segmentation applications in Table~\ref{cardiac}. Automatic cardiac segmentation is by no means a trivial task, as some parts of the cardiac borders are not obvious due to the low contrast to the surrounding tissue. Thus, the attention blocks are designed pixel-wise or consistent ($e.g.$, neighbor consistent~(\cite{liu2021deep}), interframe consistent~(\cite{ahn2021multi}), category consistent~(\cite{ding2020cab})) to identify the boundary. Moreover, the attention mechanism is also implemented in task transferring and domain transferring in the cardiac segmentation and qualification analysis tasks~\cite{ge2019k,li2020joint}.

\begin{table*}[!h]
\scriptsize
\renewcommand\arraystretch{1.25} %1.25表示1.25倍行高
\caption{An overview of non-Transformer methods for cardiac segmentation, in which Endo, Epi, ED, ES, LV, RV and MYO refers to endocardium, myocardium, end diastole, end systole, left ventricle, right ventricle and myocardium separately.}
\label{cardiac}
%\vspace{-0.5cm}
%\begin{center}
\resizebox{\linewidth}{!}{
\begin{tabular}{@{}lllllll@{}}
\toprule
Author & How & What & Datasets     & Metric                                            & \begin{tabular}[c]{@{}l@{}}Data split \\ (Train:Val:Test)\end{tabular} & Modality (Type) \\ \midrule
\cite{tong2019rianet}                 & Skip.               & Spatial                  & ACDC 2017 ~(\cite{bernard2018deep})                                                                                 & \begin{tabular}[c]{@{}l@{}}Dice:0.966 (LV ED), 0.918 (LV ES), \\ 0.948 (RV ED), 0.898 (RV ES),\\ 0.905 (MYO ED), 0.915 (MYO ES)\end{tabular}               & 100:NaN:50     & MRI (2D)                                                                          \\ \cmidrule(l){4-7 }
\cite{li2020joint}                 &  Dec                 & Spatial                  & \begin{tabular}[c]{@{}l@{}}MICCAI 2018 LA challenge\\ ~(\cite{xiong2021global})\end{tabular}                         & ACC:0.867 Dice:0.543 (Scar)                                                                                & 40:NaN:20      & MRI (3D)                                                                          \\ \cmidrule(l){4-7 }

\multirow{2}{*}{\cite{liu2021deep}} & \multirow{2}{*}{Enc} & \multirow{2}{*}{Spatial} & 1.CAMUS ~(\cite{leclerc2019deep})                                                                                   & \begin{tabular}[c]{@{}l@{}}Dice:0.951 (Endo.ED), 0.931 (Endo.ES), \\ 0.962 (Epi.ED), 0.956 (Epi.ES)\end{tabular}                                           & NM           & \multirow{2}{*}{\begin{tabular}[c]{@{}l@{}}Echocardiographic\\ (2D)\end{tabular}} \\ \cdashline{4-6}[0.8pt/5pt]
                                   &                   &                          & \begin{tabular}[c]{@{}l@{}}2.Sub-EchoNet-Dynamic\\ ~(\cite{ouyang2020video})\end{tabular}                           & \begin{tabular}[c]{@{}l@{}}Dice:0.942 (Endo.ED), 0.918 (Endo.ES),\\ 0.951 (Epi.ED), 0.943 (Epi.ES)\end{tabular}                                           & 1600:400:500 &                                                                                   \\ \cmidrule(l){4-7}
                                   
\cite{wang2022awsnet}                & Dec                  & Spatial                  & MyoPS 2020 ~(\cite{li2022myops})                                                                                & \begin{tabular}[c]{@{}l@{}}Dice:0.658 (Scar), 0.720 (Scar+Edema)\\ JC:0.535 (Scar), 0.577 (Scar+Edema)\\ HD:14.13 (Scar), 14.12 (Scar+Edema)\end{tabular} & 20:2:NaN       & CMR (2D)                                                                          \\ \bottomrule
\end{tabular}
}
\end{table*}

\textbf{Lung.} The lung-related segmentation task includes lung anatomical segmentation~(\cite{tang2019xlsor}), lung airways segmentation~(\cite{qin2020learning,tan2022segmentation}), and lesion/disease segmentation~(\cite{chen2020residual,zhang2021dense,xie2021duda,zhou2021automatic,karthik2022contour,hu2022deep,punn2020chs,yin2022sd}), most of which integrate spatial attention into the network (as shown in Table~\ref{lung}) as there are strong connections between the locations of lesions and lung. It is worth noting that adding a sub-network to offer the lungs contour maps is an efficient and common way for COVID-19 lesion segmentation~(\cite{yin2022sd,punn2020chs,xie2021duda}). Another noteworthy is that~\cite{zhang2021dense} proposed multi-layer attention specifically for COVID-19 lesion segmentation. It includes an edge attention module to learn semantic edge information from low-level features, a shape attention module with a circular shape filter to enhance attention ability for round or quasi-round pulmonary nodules, and a local attention module to learn from the adjacent region. This attention design has significantly improved the segmentation dice from 46.8\% to 56.3\%.

\begin{table*}[!h]
%\scriptsize
\renewcommand\arraystretch{1.25} %1.25表示1.25倍行高
\caption{An overview of non-Transformer methods for lung segmentation.}
\label{lung}
%\vspace{-0.5cm}
%\begin{center}
\resizebox{\linewidth}{!}{
\begin{tabular}{@{}lllllll@{}}
\toprule
Author & How & What & Datasets     & Metric                                            & \begin{tabular}[c]{@{}l@{}}Data split \\ (Train:Val:Test)\end{tabular} & Modality (Type) \\ \midrule
\multirow{2}{*}{\cite{tang2019xlsor} } &
  \multirow{2}{*}{Enc} &
  \multirow{2}{*}{Spatial} &
  1.JSRT ~(\cite{shiraishi2000development}) + Montgomery ~(\cite{jaeger2014two}) &
  Dice:0.976 &
  280:37:78 &
  \multirow{2}{*}{X-ray (3D)} \\ \cdashline{4-6}[0.8pt/5pt]
 &  &  & 2.NIH ~(\cite{tang2019xlsor})                              & Dice:0.943                       & NaN:NaN:100  &  \\ \cmidrule(l){4-7} 
\cite{qin2020learning} &
  Dec &
  Spatial &
  LIDC ~(\cite{armato2011lung}) + EXACT'09 ~(\cite{lo2012extraction})  &
  \begin{tabular}[c]{@{}l@{}} Branches detected:0.962 \\ Tree-length detected:0.907\\ TPR:0.936 FPR:0.035 Dice:0.925\end{tabular} &
  63:9:18 &
  CT (2D)\\ \cmidrule(l){4-7} 
\cite{chen2020residual} &
  Skip. &
  Spatial &
  \begin{tabular}[c]{@{}l@{}}COVID-19 CT segmentation dataset ~(\cite{medicalsegmentation.com})\end{tabular} &
  Dice:0.83 ACC:0.79 P:0.82 &
  90\%:NaN:10\% &
  CT (3D) \\ \cmidrule(l){4-7} 
\cite{zhou2021automatic} &
  Skip. &
  \begin{tabular}[c]{@{}l@{}}Spatial and \\ Channel\end{tabular} &
  \begin{tabular}[c]{@{}l@{}}COVID-19 CT segmentation datase~(\cite{medicalsegmentation.com}) \\ +  \href{http://medicalsegmentation.com/covid19/}{Segmentation dataset nr. 2} \end{tabular} &
  Dice:0.831 HD:18.8 &
  80\%:NaN:20\% &
  CT (3D) \\ \cmidrule(l){4-7} 
\cite{punn2022chs} & Skip. & Spatial &
  \begin{tabular}[c]{@{}l@{}}\href{http://medicalsegmentation.com/covid19/}{Segmentation dataset nr. 2}\\ + COVID-19-CT-Seg dataset ~(\cite{MP-COVID-19-SegBenchmark})\end{tabular}  & ACC:0.965 P:0.758 Dice:0.816 JI:0.791 & 70\%:NaN:30\% & CT (2D) \\ \cmidrule(l){4-7} 
\multirow{3}{*}{\cite{yin2022sd}} &
  \multirow{3}{*}{Enc \& Dec} &
  \multirow{3}{*}{\begin{tabular}[c]{@{}l@{}}Spatial and \\ Channel\end{tabular}} &
  \begin{tabular}[c]{@{}l@{}}1.\href{http://medicalsegmentation.com/covid19/}{COVID-19 CT segmentation dataset} \\
  + COVID-19-CT-Seg dataset ~(\cite{MP-COVID-19-SegBenchmark})\end{tabular} &
  Dice:0.8696 ACC:0.9906 JI:0.7702 &
  1373:196:391 &
  \multirow{3}{*}{CT (2D)} \\ \cdashline{4-6}[0.8pt/5pt]
 &  &  & 2.\href{http://medicalsegmentation.com/covid19/}{Segmentation dataset nr. 2}         & Dice:0.5936 ACC:0.9821 JI:0.4788 & 258:38:77      &  \\
\bottomrule
\end{tabular}
}
\end{table*}

\textbf{Kidney.}
Since renal tumors in CT or MRI images can appear similar to their parenchyma and other nearby tissues, accurately segmenting them can be challenging~(\cite{hu2019automatic,myronenko20193d,sabarinathan2019hyper,jia2022new,xuan2022dynamic}). Table~\ref{kidney} shows that spatial attention is the most commonly used attention mechanism for kidney tumor segmentation, as there is a regional inclusive relationship between the kidney and tumors~(\cite{xuan2022dynamic}) similar to the lung lesion segmentation and the information of edge is added through the spatial attention mechanism.

\begin{table*}[!h]
%\scriptsize
\renewcommand\arraystretch{1.25} %1.25表示1.25倍行高
\caption{An overview of non-Transformer methods for kidney tumor segmentation.}
\label{kidney}
%\vspace{-0.5cm}
%\begin{center}
\resizebox{\linewidth}{!}{
\begin{tabular}{@{}lllllll@{}}
\toprule
Author & How & What & Datasets     & Metric                                            & \begin{tabular}[c]{@{}l@{}}Data split \\ (Train:Val:Test)\end{tabular} & Modality (Type) \\ \midrule
\cite{myronenko20193d}   & Skip. & Spatial & KiTs ~(\cite{heller2019kits19}) & Kidney Dice:0.9742 Tumor Dice:0.8103                                                                                 & 210:NaN:90      & CT (3D) \\ \cmidrule(l){4-7}
\cite{sabarinathan2019hyper} & Dec & Spatial & KiTs ~(\cite{heller2019kits19}) & Kidney Dice:0.9535 Tumor Dice:0.8967                                                                                 & 32175:13790:NaN & CT (2D) \\ \cmidrule(l){4-7}
\cite{xuan2022dynamic}         & Skip. & Spatial & KiTs ~(\cite{heller2019kits19}) & \begin{tabular}[c]{@{}l@{}}Kidney Dice:0.961 IoU:0.926 HD:17.571\\ Tumor Dice:0.865 IoU:0.772 HD:33.839\end{tabular} & 134:34:42       & CT (3D) \\

\bottomrule
\end{tabular}
}
\end{table*}

\textbf{Liver.}
Most liver-related segmentation tasks focus on liver~(\cite{haseljic2023liver}), liver vessel~(\cite{yan2020attention, kuang2023towards}), and tumors segmentation~(\cite{jin2020ra,chen2019feature,jiang2019ahcnet,li2020attention,fan2020ma,xu2021pa,zhang2021caagp,bi2022residual,zhang2022saa}). The heterogeneous and diffusive shape, as well as the relatively small sizes of most lesions, makes the segmentation a rough task. It is noted that the spatial attention modules are usually integrated into the skip connection in those applications (shown in Table~\ref{liver}). Some researchers may prefer to perform liver segmentation before the tumor segmentation~(\cite{jin2020ra,jiang2019ahcnet}).

\begin{table*}[!h]
%\scriptsize
\renewcommand\arraystretch{1.25} %1.25表示1.25倍行高
\caption{An overview of non-Transformer methods for liver segmentation.}
\label{liver}
%\vspace{-0.5cm}
%\begin{center}
\resizebox{\linewidth}{!}{
\begin{tabular}{@{}lllllll@{}}
\toprule
Author & How & What & Datasets     & Metric                                            & \begin{tabular}[c]{@{}l@{}}Data split \\ (Train:Val:Test)\end{tabular} & Modality (Type) \\ \midrule

\cite{chen2019feature}                 &        Skip.           & Channel                                                        & LiTs ~(\cite{bilic2019liver})              & Tumor Dice:0.766                                                                                                        & 131:NaN:70    & CT (2D)                  \\ \cmidrule(l){4-7}
\multirow{3}{*}{\cite{jiang2019ahcnet}} & \multirow{3}{*}{Skip.} & \multirow{3}{*}{Spatial}                                       & 1.\begin{tabular}[c]{@{}l@{}}LiTs ~(\cite{bilic2019liver}) \end{tabular}             & NaN                                                                                                                     & 110:NaN:NaN & \multirow{3}{*}{CT (2D)} \\ \cdashline{4-6}[0.8pt/5pt]
                                     &                   &                                                                & 2.3DIRACADb ~(\cite{soler20103d})        & \begin{tabular}[c]{@{}l@{}}Liver Dice:0.959 Tumor Dice:0.734 MSD:6.271\end{tabular}                                & NaN:NaN:20  &                          \\ \cdashline{4-6}[0.8pt/5pt]
                                     &                   &                                                                & 3.Private dataset & Tumor Dice:0.591 MSD:7.538                                                                                                  & NaN:NaN:117 &                          \\ \cmidrule(l){4-7}
\multirow{2}{*}{\cite{jin2020ra}}   & \multirow{2}{*}{Skip.} & \multirow{2}{*}{Spatial}                                       & 1.LiTs ~(\cite{bilic2019liver})             & \begin{tabular}[c]{@{}l@{}}Liver Dice:0.961 JI:0.926 MSD:26.948\\ Tumor Dice:0.595 JI:0.611 MSD:6.775\end{tabular}    & 130:NaN:70  & \multirow{2}{*}{CT (3D)} \\ \cdashline{4-6}[0.8pt/5pt]
                                     &                   &                                                                & 2.3DIRACADb~(\cite{soler20103d})        & \begin{tabular}[c]{@{}l@{}}Liver: Dice:0.977 JI:0.977 MSD:18.617\\ Tumor: Dice:0.830 JI:0.744 MSD:53.324\end{tabular} & NaN:NaN:20  &                          \\ \cmidrule(l){4-7}
\cite{li2020attention}                   &         Skip.          & Spatial                                                        & LiTs ~(\cite{bilic2019liver})               & Dice:0.9815 P:0.98 R:0.99 IoU:0.9748                                                                                    & 83\%:NaN:17\% & CT (2D)                  \\ \cmidrule(l){4-7 }
\cite{fan2020ma}                     &          Dec \& Skip.         & \begin{tabular}[c]{@{}l@{}}Spatial and \\ Channel\end{tabular} & LiTs ~(\cite{bilic2019liver})               & \begin{tabular}[c]{@{}l@{}}Liver Dice:0.960 Tumor Dice:0.749 \end{tabular}                           & 90\%:10\%:70  & CT (2D)                  \\ \cmidrule(l){4-7}
\multirow{2}{*}{\cite{yan2020attention} }   & \multirow{2}{*}{Skip.} & \multirow{2}{*}{Spatial}                                       & 1.Liver vessel segmentation ~(\cite{yan2020attention})         & Dice:0.805 P:0.789 SE:0.857                                                                                            & 30:NaN:10   & \multirow{2}{*}{CT (3D)} \\ \cdashline{4-6}[0.8pt/5pt]
                                     &                   &                                                                & 2.3DIRCADb ~(\cite{soler20103d})         & Dice:0.904 P:0.990 SE:0.936                                                                                            & NaN:NaN:NaN &                          \\ 

\bottomrule
\end{tabular}
}
\end{table*}

\textbf{Polyp.}
The challenge of polyp segmentation is the diversity of size, color, and texture in the same type of polyps and the blurred boundary between a polyp and its surrounding mucosa~(\cite{fan2020pranet,tomar2021ddanet,wei2021shallow,kim2021uacanet,yeung2021focus,yang2022automatic}). Spatial attention is the mainstream to obtain spatial features with deep information, and there are some public datasets for performance comparison (in Table~\ref{polyp}). Interestingly, some researchers invariably choose to apply attention only to high-level features in parallel for different reasons.~\cite{fan2020pranet} argue that low-level feature demands more computational resource and less contributes to performance.~\cite{wei2021shallow} state that deep features are course in boundary but have a clear background and thus they utilize the clear feature to filter out the background noise in the shallow features and propose the shallow attention module.~\cite{kim2021uacanet} use the parallel axial attention on the outputs of encoder blocks and propose the uncertainty augmented context attention on the coarse-to-fine segmentation to build a bottom-up stream, which incorporates high-level semantic features for better performance.

\begin{table*}[!h]
%\scriptsize
\renewcommand\arraystretch{1.25} %1.25表示1.25倍行高
\caption{An overview of non-Transformer methods for polyp segmentation, in which mDice is the mean Dice.}
\label{polyp}
%\vspace{-0.5cm}
%\begin{center}
\resizebox{\linewidth}{!}{
\begin{tabular}{@{}lllllllll@{}}
\toprule
Author & How & What & Datasets     & Metric                                            & \begin{tabular}[c]{@{}l@{}}Data split \\ (Train:Val:Test)\end{tabular} & Modality (Type) \\ \midrule
\multirow{5}{*}{\cite{fan2020pranet}} &
  \multirow{5}{*}{Skip.} &
  \multirow{5}{*}{Spatial} &
  1.ETIS ~(\cite{silva2014toward}) &
  mDice:0.628 mIoU:0.567 MAE:0.031 &
  \multirow{5}{*}{80\%:10\%:10\%} &
  \multirow{5}{*}{Endoscopy (2D)} \\ \cdashline{4-5}[0.8pt/5pt]
 &  &  & 2.CVC-ClinicDB ~(\cite{bernal2015wm})               & mDice:0.899 mIoU:0.849 MAE:0.009 &                 &  \\ \cdashline{4-5}[0.8pt/5pt]
 &  &  & 3.CVC-ColonDB ~(\cite{bernal2012towards})                 & mDice:0.709 mIoU:0.640 MAE:0.045 &                 &  \\ \cdashline{4-5}[0.8pt/5pt]
 &  &  & 4.EndoScene ~(\cite{vazquez2017benchmark}) & mDice:0.871 mIoU:0.797 MAE:0.010 &                 &  \\ \cdashline{4-5}[0.8pt/5pt]
 &  &  & 5.Kvasir ~(\cite{jha2020kvasir})                   & mDice:0.898 mIoU:0.840 MAE:0.030 &                 &  \\ \cmidrule(l){4-7} 
\cite{tomar2021ddanet} &
  Dec &
  Channel &
  Kvasir-SEG ~(\cite{jha2020kvasir}) &
  mDice:0.8576 mIoU:0.7800 SE:0.8880 P:0.8643 &
  88\%:NaN:12\% &
  Endoscopy (2D) \\ \cmidrule(l){4-7} 
\multirow{5}{*}{\cite{wei2021shallow}} &
  \multirow{5}{*}{Dec} &
  \multirow{5}{*}{Spatial} &
  1.ETIS ~(\cite{silva2014toward}) &
  mDice:0.750 mIoU:0.654 &
  \multirow{5}{*}{80\%:10\%:10\%} &
  \multirow{5}{*}{Endoscopy (2D)} \\ \cdashline{4-5}[0.8pt/5pt]
 &  &  & 2.CVC-ClinicDB ~(\cite{bernal2015wm})       & mDice:0.916 mIoU:0.859           &                 &  \\ \cdashline{4-5}[0.8pt/5pt]
 &  &  & 3.CVC-ColonDB ~(\cite{bernal2012towards})               & mDice:0.753 mIoU:0.670           &                 &  \\ \cdashline{4-5}[0.8pt/5pt]
 &  &  & 4.EndoScene ~(\cite{vazquez2017benchmark})          & mDice:0.888 mIoU:0.815           &                 &  \\ \cdashline{4-5}[0.8pt/5pt]
 &  &  & 5.Kvasir ~(\cite{jha2020kvasir})                     & mDice:0.904 mIoU:0.847           &                 &  \\ \cmidrule(l){4-7} 
\multirow{5}{*}{\cite{kim2021uacanet} } &
  \multirow{5}{*}{Skip. \& Dec} &
  \multirow{5}{*}{Spatial} &
  1.ETIS ~(\cite{silva2014toward}) &
  mDice:0.766 mIoU:0.689 MAE:0.012 &
  NaN:NaN:100\% &
  \multirow{5}{*}{Endoscopy (2D)} \\ \cdashline{4-6}[0.8pt/5pt]
 &  &
   &
  2.CVC-ClinicDB ~(\cite{bernal2015wm}) &
  mDice:0.926 mIoU:0.880 MAE:0.006 &
  90\%:NaN:10\% &
   \\ \cdashline{4-6}[0.8pt/5pt]
 &  &  & 3.CVC-ColonDB ~(\cite{bernal2012towards})                & mDice:0.783 mIoU:0.704 MAE:0.034 & NaN:NaN:100\% &  \\ \cdashline{4-6}[0.8pt/5pt]
 &  &  & 4.CVC-300 ~(\cite{vazquez2017benchmark})                    & mDice:0.910 mIoU:0.849 MAE:0.005 & NaN:NaN:100\% &  \\ \cdashline{4-6}[0.8pt/5pt]
 &  &  & 5.Kvasir ~(\cite{jha2020kvasir})                     & mDice:0.912 mIoU:0.859 MAE:0.025 & 90\%:NaN:10\% &  \\
\bottomrule
\end{tabular}
}
\end{table*}

\textbf{Prostate.}
Developing automatic prostate segmentation remains challenging due to the missing/ambiguous boundary and inhomogeneous intensity distribution of the prostate, as well as the large variability in prostate shapes~(\cite{wang2019deep,liu2019automatic,kearney2019attention,jia2019hd,lei2020ct,xu2020asymmetrical,duran2022prostattention}). In our knowledge, the mainstream attention types is spatial attention (as shown in Table~\ref{prostate}). Moreover, the performance of these methods cannot be fairly compared since there are few publicly available dataset benchmarks. 

\begin{table*}[!h]
%\scriptsize
\renewcommand\arraystretch{1.25} %1.25表示1.25倍行高
\caption{An overview of non-Transformer methods for prostate segmentation, in which TZ is the prostatic transition zone, PZ is the peripheral zone and PB is the prostate bed.}
\label{prostate}
%\vspace{-0.5cm}
%\begin{center}
\resizebox{\linewidth}{!}{
\begin{tabular}{@{}lllllll@{}}
\toprule
Author & How & What & Datasets     & Metric                                            & \begin{tabular}[c]{@{}l@{}}Data split \\ (Train:Val:Test)\end{tabular} & Modality (Type) \\ \midrule
\cite{wang2019deep} &
  Dec &
  Spatial &
  \begin{tabular}[c]{@{}l@{}}Private dataset\end{tabular} &
  Dice:0.90 JI:0.82 HD95:8.37 P:0.90 &
  75\%:20\%:NaN &
  Ultrasound (3D) \\ \cmidrule(l){4-7} 
\multirow{2}{*}{\cite{liu2019automatic}} &
  \multirow{2}{*}{Skip.} &
  \multirow{2}{*}{Spatial} &
  1.PROSTATEX ~(\cite{litjens2014computer}) &
  Dice:0.74 (PZ), 0.86 (TZ)&
  250:NaN:63 &
  \multirow{2}{*}{MRI (2D)} \\ \cdashline{4-6}[0.8pt/5pt]
  & &
   &
  2.Private dataset &
  Dice:0.74 (PZ), 0.79 (TZ)&
  NaN:NaN:46 &
   \\ \cmidrule(l){4-7} 
\cite{kearney2019attention} &
  Skip. &
  Spatial &
  Private dataset &
  \begin{tabular}[c]{@{}l@{}}Dice:0.9002 (Prostate), 0.9312 (Bladder),\\ 0.846 (Rectum), 0.7221 (Penile bulb)\end{tabular} &
  80:20:20 &
  CT (3D) \\ \cmidrule(l){4-7} 
\cite{jia2019hd}  &
  Dec &
  Spatial &
  PROMISE12 ~(\cite{litjens2014evaluation}) &
  Dice:0.9135 HD95:3.93 Score:90.34 &
  50:NaN:30 &
  MRI (3D) \\ \cmidrule(l){4-7} 
\cite{lei2020ct} &
  Skip. &
  Spatial &
  Private dataset &
  Dice:0.91 HD:4.57 MSD:0.62 &
  49:NaN:50 &
  CT (3D) \\ \cmidrule(l){4-7} 
\cite{xu2020asymmetrical}  &
  Skip. &
  Spatial &
  Private dataset &
  \begin{tabular}[c]{@{}l@{}}Dice:0.7567 (PB), 0.8840 (Bladder), 0.8035 (Rectum) \\ ASD:0.42 (PB), 1.47 (Bladder), 2.60 (Rectum)\end{tabular} &
  60\%:20\%:20\% &
  CT (2D) \\
\bottomrule
\end{tabular}
}
\end{table*}

\textbf{Retinal.}
Retinal segmentation contains the iris segmentation~(\cite{lian2018attention}), disc and cups segmentation~(\cite{jiang2019jointrcnn,zhang2019net,bhatkalkar2020improving}), and the retinal vessel segmentation~(\cite{mou2019cs,zhang2019attention,li2019connection,luo2019micro,wu2019u,wang2019dual,hu2019s,li2020lightweight,guo2021sa,lyu2020attention,lv2020attention,guo2020residual,li2020accurate,tong2021sat,guo2021channel,li2021bseresu,wu2021encoding,jiang2021bi,wang2020hard,wang2022attention,liu2022multiscale,yang2022dcu})
, while the last one is the majority. We observe that the spatial-wise attention module is mostly inserted into the skip connection in the retinal vessel segmentation (as shown in Table~\ref{retinal}) as the characteristic spatial information of different scales can help to better extract vessels due to the blurring of vessel boundary and the reflection of vessel centerline in fundus images~(\cite{liu2022multiscale}). Especially, hard attention is first introduced into retinal vessel segmentation by~\cite{wang2020hard} in HAnet, which is composed of one encoder and three decoder sub-networks. Specifically, a basic decoder is expected to yield a coarse vessel segmentation result and provide a probabilistic map to automatically determine which region is “easy” or “hard” to segment. The attention mechanism is integrated into the “hard” decoder branch to effectively reinforce vessel features. HAnet achieves the highest segmentation accuracy and area under the receiver operating characteristic curve (AUC) on the public fundus datasets. 

\begin{table*}[!h]
%\scriptsize
\renewcommand\arraystretch{1.25} %1.25表示1.25倍行高
\caption{An overview of non-Transformer methods for retinal vessel segmentation.}
\label{retinal}
%\vspace{-0.5cm}
%\begin{center}
\resizebox{\linewidth}{!}{
\begin{tabular}{@{}lllllll@{}}
\toprule
Author & How & What & Datasets     & Metric                                            & \begin{tabular}[c]{@{}l@{}}Data split \\ (Train:Val:Test)\end{tabular} & Modality (Type) \\ \midrule
\multirow{3}{*}{\cite{zhang2019attention} } &
  \multirow{3}{*}{Skip.} &
  \multirow{3}{*}{Spatial} &
  1.DRIVE ~(\cite{staal2004ridge}) &
  ACC:0.9692 AUC:0.9856 IoU:0.6965 &
  20:NaN:20 &
  \multirow{3}{*}{Retinal fundus (2D)} \\ \cdashline{4-6}[0.8pt/5pt]
 &  &
   &
  2.CHASEDB1 ~(\cite{owen2009measuring}) &
  ACC:0.9743 AUC:0.9863 IoU:0.6669 &
  NaN:NaN:100\% &
   \\ \cdashline{4-6}[0.8pt/5pt]
 &  &
   &
  3.ORIGA ~(\cite{zhang2010origa}) &
  OE:0.061 (Diac), 0.212 (Cup), 0.137 (Total)&
  325:NaN:325 &
   \\ \cmidrule(l){4-7} 
\multirow{3}{*}{\cite{li2019connection}} &
  \multirow{3}{*}{Skip.} &
  \multirow{3}{*}{Spatial} &
  1.DRIVE ~(\cite{staal2004ridge}) &
  AUC:0.9807 ACC:0.9560 &
  20:NaN:20 &
  \multirow{3}{*}{Retinal fundus (2D)} \\ \cdashline{4-6}[0.8pt/5pt]
 &  &
   &
  2.STARE ~(\cite{hoover2000locating}) &
  AUC:0.9834 ACC:0.9673 &
  10:NaN:10 &
   \\ \cdashline{4-6}[0.8pt/5pt]
 &  &
   &
  3.HRF ~(\cite{budai2013robust}) &
  AUC:0.9867 &
  15:NaN:30 &
   \\ \cmidrule(l){4-7} 
\multirow{5}{*}{\cite{mou2019cs} } &
  \multirow{5}{*}{Skip.} &
  \multirow{5}{*}{\begin{tabular}[c]{@{}l@{}}Spatial and \\ Channel\end{tabular}} &
  1.DRIVE ~(\cite{staal2004ridge}) &
  ACC:0.9632 SE:0.8215 AUC:0.9825 &
  80\%:NaN:20\% &
  Retinal fundus (2D) \\ \cdashline{4-7}[0.8pt/5pt]
 &  &
   &
  2.STARE ~(\cite{hoover2000locating}) &
  ACC:0.9752 SE:0.8816 AUC:0.9932 &
  75\%:NaN:25\% &
  Retinal fundus (2D) \\ \cdashline{4-7}[0.8pt/5pt]
 &  &
   &
  3.Private dataset &
  ACC:0.9183 SE:0.8631 AUC:0.9453 &
  80\%:NaN:20\% &
  OCT-A (2D) \\ \cdashline{4-7}[0.8pt/5pt]
 &  &
   &
  4.CMM-1 ~(\cite{imedcorn}) &
  SE:0.8415 FDR:0.2521 &
  80\%:NaN:20\% &
  Microscopy (2D) \\ \cdashline{4-7}[0.8pt/5pt]
 &  &
   &
  5.CMM-2 ~(\cite{imedcorn}) &
  SE:0.8345 FDR:0.2591 &
  80\%:NaN:20\% &
  Microscopy (2D) \\ \cmidrule(l){4-7} 
\multirow{5}{*}{\cite{li2020lightweight} } &
  \multirow{5}{*}{Dec} &
  \multirow{5}{*}{Spatial} &
  1.DRIVE ~(\cite{staal2004ridge}) &
  ACC:0.9568 AUC:0.9806 &
  20:NaN:20 &
  \multirow{5}{*}{Retinal fundus (2D)} \\ \cdashline{4-6}[0.8pt/5pt]
 &  &
   &
  2.STARE ~(\cite{hoover2000locating}) &
  ACC:0.9678 AUC:0.9678 &
  19:NaN:1 &
   \\ \cdashline{4-6}[0.8pt/5pt]
 &  &
   &
  3.CHASEDB1 ~(\cite{owen2009measuring}) &
  ACC:0.9635 AUC:0.9702 &
  20:NaN:8 &
   \\ \cdashline{4-6}[0.8pt/5pt]
 &  &
   &
  4.IOSTAR ~(\cite{zhang2016robust}) &
  ACC:0.9544 AUC:0.9623 &
  20:NaN:10 &
   \\ \cdashline{4-6}[0.8pt/5pt]
 &  &
   &
  5.RC-SLO ~(\cite{abbasi2015biologically})&
  AUC:0.9696 AUC:0.8119 &
  30:NaN:10 &
   \\ \cmidrule(l){4-7} 
\multirow{2}{*}{\cite{guo2021sa} } &
  \multirow{2}{*}{Enc.} &
  \multirow{2}{*}{Spatial} &
  1.DRIVE ~(\cite{staal2004ridge}) &
  ACC:0.9698 AUC:0.9864 &
  20:NaN:20 &
  \multirow{2}{*}{Retinal fundus (2D)} \\ \cdashline{4-6}[0.8pt/5pt]
 &  &
   &
  2.CHASEDB1 ~(\cite{owen2009measuring}) &
  ACC:0.9755 AUC:0.9905 &
  20:NaN:8 &
   \\ \cmidrule(l){4-7} 
\multirow{3}{*}{\cite{lv2020attention}} &
  \multirow{3}{*}{Enc.} &
  \multirow{3}{*}{Spatial} &
  1.DRIVE ~(\cite{staal2004ridge}) &
  ACC:0.9558 AUC:0.9847 &
  20:NaN:20 &
  \multirow{3}{*}{Retinal fundus (2D)} \\ \cdashline{4-6}[0.8pt/5pt]
 &  &
   &
  2.STARE ~(\cite{hoover2000locating}) &
  ACC:0.9640 AUC:0.9824 &
  10:NaN:10 &
   \\ \cdashline{4-6}[0.8pt/5pt]
 &  &
   &
  3.CHASEDB1 ~(\cite{owen2009measuring}) &
  ACC:0.9608 AUC:0.9865 &
  20:NaN:8 &
   \\ \cmidrule(l){4-7} 
\multirow{3}{*}{\cite{li2020accurate} } &
  \multirow{3}{*}{Dec} &
  \multirow{3}{*}{Spatial} &
  1.DRIVE ~(\cite{staal2004ridge}) &
  ACC:0.9769 AUC:0.9895 &
  20:NaN:20 &
  \multirow{3}{*}{Retinal fundus (2D)} \\ \cdashline{4-6}[0.8pt/5pt]
 &  &
   &
  2.STARE ~(\cite{hoover2000locating}) &
  ACC:0.9797 AUC:0.9924 &
  10:NaN:10 &
   \\ \cdashline{4-6}[0.8pt/5pt]
 &  &
   &
  3.CHASEDB1 ~(\cite{owen2009measuring}) &
  ACC:0.9803 AUC:0.9912 &
  14:NaN:14 &
   \\ \cmidrule(l){4-7} 
\multirow{3}{*}{\cite{guo2021channel} } &
  \multirow{3}{*}{Skip.} &
  \multirow{3}{*}{Channel} &
  1.DRIVE ~(\cite{staal2004ridge}) &
  ACC:0.9699 AUC:0.9852 &
  20:NaN:20 &
  \multirow{3}{*}{Retinal fundus (2D)} \\ \cdashline{4-6}[0.8pt/5pt]
 &  &
   &
  2.STARE ~(\cite{hoover2000locating}) &
  ACC:0.9743 AUC:0.9911 &
  15:NaN:15 &
   \\ \cdashline{4-6}[0.8pt/5pt]
 &  &
   &
  3.CHASEDB1 ~(\cite{owen2009measuring}) &
  ACC:0.9751 AUC:0.9898 &
  20:NaN:8 &
   \\ \cmidrule(l){4-7} 
\multirow{4}{*}{\cite{wang2020hard}} &
  \multirow{4}{*}{Skip.} &
  \multirow{4}{*}{Spatial} &
  1.DRIVE ~(\cite{staal2004ridge}) &
  AUC:0.9823 ACC:0.9581 &
  20:NaN:20 &
  \multirow{4}{*}{Retinal fundus (2D)} \\ \cdashline{4-6}[0.8pt/5pt]
 &  &
   &
  2.STARE ~(\cite{hoover2000locating}) &
  AUC:0.9881 ACC:0.9673 &
  19:NaN:1 &
   \\ \cdashline{4-6}[0.8pt/5pt]
 &  &
   &
  3.CHASEDB1 ~(\cite{owen2009measuring}) &
  AUC:0.9871 ACC:0.9670 &
  20:NaN:8 &
   \\ \cdashline{4-6}[0.8pt/5pt]
 &  &
   &
  4.HRF ~(\cite{budai2013robust}) &
  AUC:0.9837 ACC:0.9654 &
  15:NaN:30 &
   \\ \cmidrule(l){4-7} 
\multirow{2}{*}{\cite{liu2022multiscale} } &
  \multirow{2}{*}{Skip.} &
  \multirow{2}{*}{\begin{tabular}[c]{@{}l@{}}Spatial and \\ Channel\end{tabular}} &
  1.DRIVE ~(\cite{staal2004ridge}) &
  ACC:0.9699 &
  20:NaN:20 &
  \multirow{2}{*}{Retinal fundus (2D)} \\ \cdashline{4-6}[0.8pt/5pt]
 &  &
   &
  2.CHASEDB1 ~(\cite{owen2009measuring}) &
  ACC:0.9751 &
  14:NaN:14 &
   \\
\bottomrule
\end{tabular}
}
\end{table*}

\textbf{Skin Lesion.}
Due to the influence of color, boundaries, and shapes of melanoma as well as various artifacts, the segmentation of the skin lesion area is still a challenging problem~(\cite{kaul2019focusnet,wei2019attention,singh2019fca,wu2020automated,tong2021ascu,ren2022serial,arora2021automated,wang2022skin}). 
 Here, we observe that spatial attention is the most commonly used mechanism in the skip connection for skin lesion segmentation, as shown in Table~\ref{skin lesion}. This may be due to the importance of multi-scale information in handling skin lesions of varying sizes and shapes. By incorporating spatial attention in the skip connection, the down-sampling path ensures the maximum flow of multi-scale information between layers, allowing for more accurate and effective segmentation of skin lesions, regardless of their size or shape.
Besides, multi-resolution inputs are adopted to learn better specific and discriminative features in these applications~(\cite{singh2019fca,wu2020automated,wang2022skin}). Based on that,~\cite{wu2020automated} propose an adaptive dual attention module, integrating global context and pixel-wise correlation, with different dilation rates for different sizes of skin lesions, and the model performs best on both ISBI2017~(\cite{codella2018skin}) and ISIC2018 datasets~(\cite{tschandl2018ham10000}).

\begin{table*}[!h]
\scriptsize
\renewcommand\arraystretch{1.25} %1.25表示1.25倍行高
\caption{An overview of non-Transformer methods for skin lesion segmentation.}
\label{skin lesion}
%\vspace{-0.5cm}
%\begin{center}
\resizebox{\linewidth}{!}{
\begin{tabular}{@{}lllllll@{}}
\toprule
Author & How & What & Datasets     & Metric                                            & \begin{tabular}[c]{@{}l@{}}Data split \\ (Train:Val:Test)\end{tabular} & Modality (Type) \\ \midrule
\multirow{3}{*}{\cite{wei2019attention} } &
  \multirow{3}{*}{Skip.} &
  \multirow{3}{*}{Spatial} &
  1.ISBI 2016 ~(\cite{gutman2016skin}) &
  ACC:0.9683 Dice:0.9536 JI:0.9142&
  900:NaN:379 &
  \multirow{3}{*}{Dermoscopy(2D)} \\ \cdashline{4-6}[0.8pt/5pt]
 &  &  & 2.ISBI 2017 ~(\cite{codella2018skin}) & ACC:0.9329 Dice:0.8786 JI:0.8045  & 2000:150:600   &  \\ \cdashline{4-6}[0.8pt/5pt]
 &  &  & 3.PH2 ~(\cite{mendoncca2013ph})      & Divergence Value:8.23         & NaN:NaN:200  &  \\ \cmidrule(l){4-7} 
\multirow{2}{*}{\cite{wu2020automated}} &
  \multirow{2}{*}{Skip.} &
  \multirow{2}{*}{\begin{tabular}[c]{@{}l@{}}Spatial and \\ Channel\end{tabular}} &
  1.ISBI 2017 ~(\cite{codella2018skin}) &
  ACC:0.9570 Dice:0.8969 JI:0.8255 &
  2000:150:600 &
  \multirow{2}{*}{Dermoscopy (2D)} \\ \cdashline{4-6}[0.8pt/5pt]
 &  &  & 2.ISIC 2018 ~(\cite{tschandl2018ham10000}) & ACC:0.9470 Dice:0.9080 JI:0.8440 & 2000:594:NaN &  \\ \cmidrule(l){4-7} 
\cite{arora2021automated}  &
  Skip. &
  Spatial &
  ISIC 2018 ~(\cite{tschandl2018ham10000}) &
  ACC:0.95 Dice:0.91 JI:0.83 &
  2000:594:NaN &
  Dermoscopy (2D) \\ \cmidrule(l){4-7} 
\multirow{3}{*}{\cite{tong2021ascu} } &
  \multirow{3}{*}{Dec} &
  \multirow{3}{*}{\begin{tabular}[c]{@{}l@{}}Spatial and \\ Channel\end{tabular}} &
  1.ISIC 2016 ~(\cite{gutman2016skin}) &
  ACC:0.9540 JI:0.8450 &
  2000:150:600 &
  \multirow{3}{*}{Dermoscopy (2D)} \\ \cdashline{4-6}[0.8pt/5pt]
 &  &  & 2.ISIC 2017 ~(\cite{codella2018skin})  & ACC:0.9260 JI:0.7420           & 900:379:NaN  &  \\ \cdashline{4-6}[0.8pt/5pt]
 &  &  & 3.PH2 ~(\cite{mendoncca2013ph})       & ACC:0.9430 JI:0.8420           & NaN:NaN:200  &  \\
\bottomrule
\end{tabular}
}
\end{table*}

\textbf{Surgical Instrument.}
Since robot-assisted surgery has gained increasing popularity, segmentation for tracking instruments has attracted more attention. However, the geometry change due to the pose changing, the partial occlusion caused by the narrow field of view, the specular reflection, and the serious class imbalance obstacles the automatic surgical instrument segmentation.

In the early works, the attention mechanism is introduced into the ResNet with channel/spatial attention~\cite{ni2019rasnet,ni2019raunet}. Then the complementary attention is integrated into the network~\cite{ni2022surginet}, in which a position attention block and a channel attention block are combined into a double attention module. Moreover, the inherent temporal clues from the instrument motion are also taken into account~(\cite{jin2019incorporating}). Specifically, the prediction mask of the previous frame can be regarded as temporal prior to the instrument’s current location and shape, thus integrated into the segmentation network as an initial attention gate signal. 
Existing medical surgical instrument dataset includes Cata7~(\cite{ni2019raunet}), CataIS~(\cite{ni2019rasnet}), EndoVis 2017~(\cite{allan20192017}), and ROBUST-MIS dataset~(\cite{maier2021heidelberg}), in which Cata7 and CataIS are constructed by~\cite{ni2019rasnet,ni2019raunet} for cataract surgical instrument segmentation from Beijing Tongren Hospital and will be public soon. EndoVis 2017~(\cite{allan20192017}) is from the MICCAI Endovis Challenge 2017, which is based on endoscopic surgery with 3000 images and ROBUST-MIS dataset is comprised of a total of 10,040 annotated video frames from 30 minimally invasive daily-routine surgical procedures.

\textbf{Conclusion.}
In short, attention mechanisms have been applied to various aspects of medical image segmentation, and the number of papers depends on the number of publicly available datasets in this domain. We also observe that spatial attention is more popular than channel attention, this may be because edge information is of significance for blurring boundaries (polyp segmentation, prostate segmentation and skin lesion segmentation. etc.) and there is a strong connection between lesions/tumors and the corresponding organs (e.g., the lung lesion segmentation, kidney tumor segmentation, liver tumor segmentation, etc.). What needs to be emphasized is that very few researchers design the attention module for organ-specific.

% ####################################################

\section{Transformer in medical Segmentation}
\label{sec4}
To provide a comprehensive understanding of Transformer-based methods in medical segmentation, we have structured our survey to follow the format of the Non-Transformer part. We first introduce the core concept in Transformer (what to use), followed by the mainstream network architecture (how to use), and the specific application tasks (where to use).

\subsection{What to Use}
The basic Transformer layer~(\cite{vaswani2017attention}) comprises two main sub-layers: the multi-head self-attention and feedforward layers. 
Self-attention mechanism employs scaled dot-product attention to model interactions between all elements of a sequence, given by equation (2). 
% 
% The input vector is first transformed into three distinct matrix representations: queries $Q \in \mathbb{R}^{n\times d_{q}}$, keys $K \in \mathbb{R}^{n\times d_{k}}$, and values $V \in \mathbb{R}^{n\times d_{v}}$, all with dimensions $d_{q}=d_{k}=d_{v}=d_{model}$. The calculation formula is given by
% \begin{equation}
%     {\rm Attention}(Q,K,V)={\rm softmax}(\frac{QK^{T}}{\sqrt{d_{k}}})V
% \end{equation}
% where
% % $n$ denotes the length of sequence, 
% $QK^T$ computes the relevance score between different entities, $d_{k}$ is the scaling factor, softmax operation translates the score into probability and multiplying with $V$ is to obtain the weighted matrix. Vectors with larger probabilities receive higher focus from the following layers.
% 
The multi-head attention mechanism captures complex relationships between entities by allowing attention layers to focus on different representation subspaces. Given an input vector and the number of heads $h$, the input vector is transformed into three separate groups of vectors, each containing $h$ vectors. The following equation can describe this process:
\begin{equation}
\begin{split}
    &Q=\left\{Q_{i} \right\}_{i=1}^{h}, K=\left\{K_{i} \right\}_{i=1}^{h}, V=\left\{V_{i} \right\}_{i=1}^{h} \\
    &{\rm head}_{i}={\rm Attention}(Q_{i},K_{i},V_{i}) \\
    &{\rm MultiHead}(Q,K,V)=Concat({\rm head}_{1},\cdots,{\rm head}_{h})W^{O} 
\end{split}
\end{equation}
% where $Q$, $K$, $V$ represent the concatenation of $\left\{Q_{i} \right\}_{i=1}^{h}$, $\left\{K_{i} \right\}_{i=1}^{h}$, $\left\{V_{i} \right\}_{i=1}^{h}$, which is the subspace representation respectively. 
where $W^{O}\in\mathbb{R}^{d_{model}\times d_{model}}$ is the linear projection matrix.

The feedforward layer applies a two-layer feedforward neural network to the attended input. It consists of two linear transformations with a ReLU activation function in between. This allows the model to transform the attended input into a different space and learn more complex representations. In addition to these two sub-layers, the Transformer layer includes residual connections and layer normalization. The residual connections allow the model to learn identity mapping and mitigate the vanishing gradient problem. Layer normalization helps to stabilize the training process by normalizing the inputs to each layer.

The Transformer layer can be stacked multiple times to form the encoder and decoder networks in the Transformer architecture. The encoder takes an input sequence and produces a sequence of hidden representations, while the decoder takes a sequence of hidden representations and produces an output sequence.
% The Transformer is an encoder-decoder network at a high level, which use stacked blocks including a self-attention sub-layer and a fully connected sub-layer for both encoder and decoder. Specifically, an additional masked self-attention sub-layer is added into the decoder block to prevent positions from attending to subsequent positions.

Some successful Transformer-based works have been proposed in computer vision. 
\textbf{Vision Transformer (ViT},~\cite{dosovitskiy2020image}) is the first model to introduce the Transformer encoder into computer vision for image classification, which applies the Transformer model on a sequence of image patches flattened as vectors directly with a 1D learnable positional encoding. ViT inserts a learned \emph{[class]} embedding whose state at the output of the Transformer encoder serves as a representation to perform classification. ViT was shown to achieve state-of-the-art performance on the ImageNet benchmark dataset, demonstrating the effectiveness of the Transformer architecture for computer vision tasks.
\textbf{DEtection TRansformer (DETR},~\cite{carion2020end}) is a Transformer-based model for object detection that treats object detection as a direct set prediction problem. It uses an encoder-decoder architecture based on the Transformer to predict all objects simultaneously. The Transformer module outputs \emph{N} objects' embeddings in parallel after a CNN backbone for \emph{N} final predictions (including box coordinates and class labels). \textbf{Deformable DETR}~(\cite{zhu2020deformable}) is a variation of DETR that proposes the deformable attention module to mitigate the high computational cost issues in DETR. Deformable attention attends to a sparse set of elements from the whole feature map regardless of all elements.
\textbf{SEgmentation TRansformer (SETR},~\cite{zheng2021rethinking}) attempts to migrate the Transformer to image segmentation tasks, with a pure Transformer encoder and a CNN-based decoder. Three fashions of the encoder are employed for pixel-wise classification: naive up-sampling, progressive up-sampling, and multi-level feature aggregation for exploration.
\textbf{Pyramid Vision Transformer (PVT},~\cite{wang2021pyramid}) introduces a progressive shrinking pyramid with pure Transformer block and adopts a spatial-reduction attention layer to reduce the computational costs in dense prediction tasks.
Unlike the above transformer-based models that operate on fixed-sized image patches, \textbf{Swin Transformer}~(\cite{liu2021swin}) introduces the shifted window operation to efficiently capture global image features while preserving fine-grained spatial information. Swin Transformer also incorporates several novel techniques, such as shifted window attention and local feature aggregation. With its hierarchical design, Swin Transformer has achieved impressive results on several benchmark datasets, demonstrating its effectiveness for various vision tasks.

In medical image segmentation tasks, Transformer and its improved versions ($e.g.$, Swin Transformer) are usually introduced into the U-Net architecture as a plug-in module or the basic block.  

\subsection{How to Use}

% Hybrid encoder —— CNN decoder  （neck，skip，encoder）
% Pure Trans encoder —— CNN decoder
% Pure Trans encoder —— Trans decoder  

\begin{figure}[!h]
    \centering
    \centerline{\includegraphics[width=\linewidth]{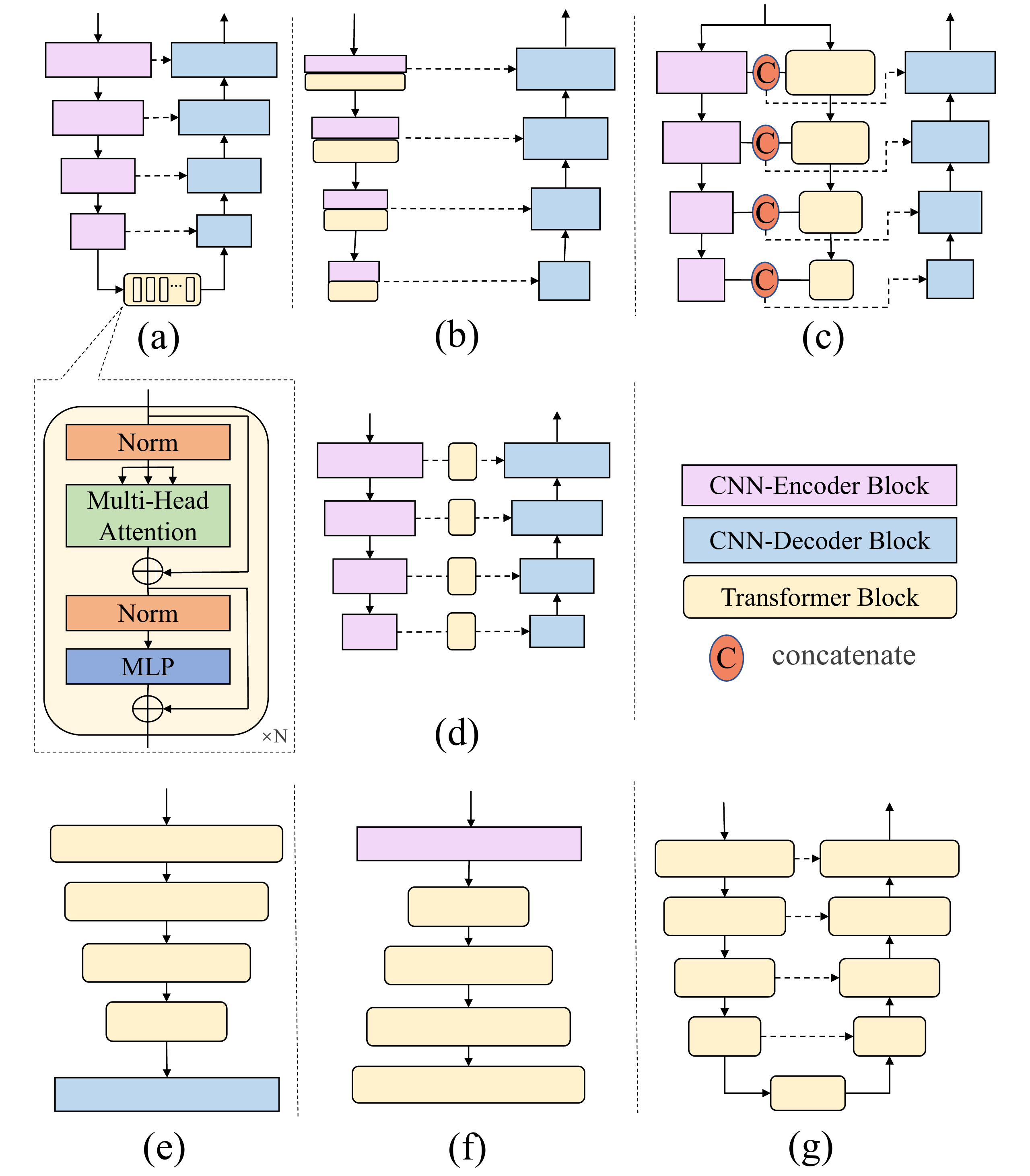}}
    \caption{Comparison of different Transformer-based architectures for medical image segmentation. The first two rows represent Hybrid encoder + CNN decoder cases: (a) Transformer used as the bottleneck, (b) Transformer combined with CNN in serial as encoder, (c) Transformer combined with CNN in parallel as a dual-path encoder, and (d) Transformer in the skip connection. The last row represents cases where Transformer is regarded as the main body in the encoder/decoder: (e) Transformer encoder + CNN decoder, where Transformer stacks are in pyramid style in the encoder, (f) CNN encoder + Transformer decoder, with Transformer stacks in the decoder, and (g) Transformer encoder + Transformer decoder. Please note that the decoder in (e) and the encoder in (f) are simplified due to limited space.}
    \label{Transformer-archit in segmentation}
\end{figure}

Despite Transformer's ability to model the global contextual dependency, the self-attention induces missing inductive bias of locality~(\cite{dosovitskiy2020image}). Meanwhile, convolution block with locality advances to deal with these features with preferable inductive bias~(\cite{simoncelli2001natural}) ($e.g.$, translation invariance~(\cite{scherer2010evaluation})). Thus Transformer-based methods are more likely to combine with CNN by leveraging the locality of CNNs and the long-range dependency character of the Transformer for medical image segmentation with limited data.
Here, we observe that Transformer-based models can be categorized by the main architecture design as shown in Fig~\ref{Transformer-archit in segmentation}, consisting of Hybrid encoder + CNN decoder, Pure Transformer encoder + CNN decoder, CNN encoder + Pure Transformer decoder, and Transformer encoder + Transformer decoder.

\subsubsection{Hybrid encoder + CNN decoder}
U-Net and its modified versions based on CNN have been widely used in medical image segmentation due to their efficient and precise feature extraction capabilities. To further improve the performance of U-Net and its variants, researchers have explored integrating the Transformer into these architectures to improve the long-term modeling ability. This is typically done by adding Transformer in the encoder to form a hybrid encoder leveraging both benefits of Transformer and CNN. Here we categorize these methods into four groups according to Transformer's location and proportion in the U-Net style network.

\textbf{Transformer as the bottleneck.}
A most natural and source-saving form is introducing the Transformer as the bottleneck in the encoder (shown in Fig\ref{Transformer-archit in segmentation} (a)).~\cite{chen2021transunet} propose TransUnet, which integrates ViT into the U-Net architecture to capture long-term dependencies in the CNN. TransUnet has shown promising results on 2D medical images and has inspired further research in integrating Transformers into U-Net for medical image segmentation.
For multi-task learning,~\cite{cheng2022fully} follow the TransUNet design for glioma segmentation and combine the features from CNN and Transformer for isocitrate dehydrogenase (IDH) genotyping.~\cite{zhang2021multi} implement a vanilla Transformer encoder-decoder as a bottleneck module within their approach. After the U-Net processing, the resulting feature is divided into two branches: body and edge. This division promotes local consistency and provides edge position information.~\cite{wang2021boundary} also include the Transformer encoder-decoder as a bottleneck but introduce boundary-wise prior knowledge for guidance, enhancing the model's performance in tasks that require a focus on boundaries.
To reduce the information loss during the upsampling operations,~\cite{yang2022transnunet} add CBAM~(\cite{woo2018cbam}) in upsampling operations to capture the region of interest.~\cite{chen2021transattunet} design a novel Transformer module as the bottleneck. It consists of a Transformer Self-Attention module to jointly attend to semantic information from the global representation subspace and a Global Spatial Attention module to selectively aggregate global context to the learned features.~\cite{chang2021transclaw} insert the Transformer block in the bottleneck and introduce multi-scale properties by upsampling the Transformer's feature to different scales and concatenating them with each layer feature from the CNN encoder. LeviT-UNet~(\cite{xu2021levit}) uses LeViT~(\cite{graham2021levit}) instead of the vanilla Transformer to reduce the computational cost. 
Meanwhile,~\cite{guo2021transformer} and~\cite{yan2022after} try to take 3D information into account in the 2D network with the help of the Transformer bottleneck block, specifically, the former encodes dependencies between slices along the z-axis for the 3D anisotropy problem and the latter proposes an axial Transformer to leverage intra- and inter-slice contextual information.
TransBTS~(\cite{wang2021transbts}) improve a 3D version of TransUNet~(\cite{chen2021transunet}). After that, TransBTSV2 version~(\cite{li2022transbtsv2}) inserts an expansion module to expand the Transformer width. To boost the performance of TransBTS,~\cite{jia2021bitr} prefer to add another Transformer layer into the penultimate layer to get dense information, while~\cite{dobko2021combining} add SE blocks~(\cite{hu2018squeeze}) to each layer of the encoder block in TransBTS~(\cite{wang2021transbts}), meanwhile, the positional encoding is replaced with a learnable MLP block.

Cross-scale dependency and consistency are crucial in segmenting objects, such as lesions, that experience significant size changes. To address this challenge, researchers have proposed various methods that involve concatenating encoded features of different scales and sending them to a Transformer block within the bottleneck.
CoTr~(\cite{xie2021cotr}) applies an efficient deformable Transformer on the concatenated multi-layer feature to model multi-scale contextual features, paying attention to only a small set of key positions to reduce the computational and spatial complexities.~\cite{wang2021mstganet} design a novel Transformer module and place it at the top of the encoder path, aiming to capture multi-scale non-local features with long-range dependencies from different layers of the encoder.
UCTransUNet~(\cite{wang2021uctransnet}) designs a channel-wise cross fusion Transformer to capture local-channel interaction and solve the inconsistent semantic level problem, which replaces the simple skip connection. In this method, the concatenated features from the encoder act as the key and value, and the feature from each layer is the query in the cross fusion Transformer.~\cite{ji2021multi} embed the multi-scale convolutional features as a sequence of tokens and perform Transformer self-attention and cross-attention sequentially to capture the cross-scale dependencies. Additionally, they introduce a learnable proxy embedding to model semantic relationships and use it as the query in the cross-attention module.

\textbf{CNN-Transformer in Serial as Encoder.}
Incorporating Transformers only as a bottleneck may not effectively capture long-term dependencies, thus researchers have introduced Transformer modules into more network stages, enhancing complementary benefits and improving performance.

As shown in Fig~\ref{Transformer-archit in segmentation} (b), Transformers can be introduced after each convolutional layer to build long-range dependency on features of various scales. 
UTNet~(\cite{gao2021utnet}) adds a Transformer layer after each convolutional operation in the encoder and designs an efficient self-attention mechanism along with relative position encoding to reduce complexity. 
SpecTr~(\cite{yun2021spectr}) employs a similar architecture, but with a sparsity scheme and spectral normalization strategy in the Transformer block for hyperspectral pathology image segmentation. 
\cite{wang2021ccut} propose a Transformer with a SE operation at each scale and design a CCA block consisting of GCNet~(\cite{cao2019gcnet}) and ECA-Net~(\cite{Wang_2020_CVPR}) to optimize the model by combining both spatial and channel attention in the skip connection. RTNet~(\cite{huang2022rtnet}) introduces a global Transformer and a relation Transformer to capture inter- and intra-information between vessel and lesion features for retinopathy multi-lesion segmentation. 
Additionally,~\cite{wang2021multi} integrate CNN-Transformer information through concatenation, rather than operating in sequence, and propose a 3D image position embedding that allows the local neighbourhood to facilitate role in global attention within multi-head self-attention.

\textbf{CNN-Transformer in Dual-Path as Encoder.}
Some works adopt a dual-path encoder, combining Transformers and CNNs, to simultaneously learn global and local dependencies, as shown in Fig~\ref{Transformer-archit in segmentation} (c).~\cite{sun2021hybridctrm} propose a dual-path CNN-Transformer encoder and concatenate the final outputs of both at the network's bottom as the decoder's input.~\cite{wu2022fat} extend this model by adding a SE block~(\cite{hu2018squeeze}) on the fused feature for channel information.~\cite{sha2021transformer} also employ a dual-path encoder and concatenate feature maps from both encoders at each layer.  
To further combine the information from each layer,~\cite{zhang2021pyramid} design three axial Transformer branches in different input scales and a CNN branch, integrating them at three levels to grasp associations of diverse scales within the image. 
Similarly,~\cite{zhang2021transfuse} report Transfuse, where features with the same resolution are fused through self-attention and bilinear Hadamard product, acting as signal gates between a CNN and a Transformer branch. 
Besides,~\cite{reza2022contextual} present a different fusion method, adding image-level contextual representation and regional importance coefficients from Transformers to CNNs for spatial normalization.

% Here, it is obvious that the dual-path framework is time-saving while the serial framework is space-saving. However, there is still no experiment to discuss the advantages, disadvantages, time efficiency, local/global information interaction, and the information loss problem of these two structures.
Indeed, the dual-path framework offers time-saving benefits, while the serial framework is more space-saving. However, there has not been a comprehensive experiment or study that thoroughly compares these two structures in terms of their advantages, disadvantages, time efficiency, local/global information interaction, and potential information loss. Such a study would provide valuable insights into the most effective approach for specific applications and guide future research in the medical image segmentation domain.

\textbf{Transformer in the Skip Connection.}
Some approaches utilize Transformers in the skip connection to bring global information into CNN architectures, as shown in Fig~\ref{Transformer-archit in segmentation} (d).~\cite{yu2022vision} add channel attention vision transformer (CAViT) in the skip connection, combining Transformers and ECA~(\cite{Wang_2020_CVPR}) blocks to leverage both channel attention and self-attention. They also introduce a deep adaptive gamma correlation to improve retinal segmentation.~\cite{petit2021u} apply a multi-head cross-attention in the skip connection to filter non-semantic richness features as well as a multi-head self-attention at the top of U-Net encoder to leverage global interactions between semantic features.~\cite{you2022class} build a GAN to enhance performances, incorporating a class-aware Transformer module in the skip connection to learn regions of interest in the generator progressively. They also devise a ResNet-Transformer discriminator for improved performance.

\subsubsection{Pure Transformer encoder + CNN decoder}
To take advantage of the U-Net architecture design, Transformer layers can be stacked in a pyramid fashion inspired by PVT~(\cite{wang2021pyramid}), as shown in the last row of Fig \ref{Transformer-archit in segmentation}. The building blocks include multilayer perceptrons (MLP), multi-head self-attention layers (MSA), and appropriate down-sampling layers. Thus they can learn hierarchical object concepts at different resolutions.
% However, Transformer offers high accuracy while coming at the cost of very high computational complexity hindering their development in this application manner.

As an example, those methods introduce Transformer only into the encoder to replace the convolution block.~\cite{karimi2021convolution} and UNETR~(\cite{hatamizadeh2022unetr}) applies ViT-like Transformer backbone as a substitute for the convolution block in the encoder part in a U-Net-like structure. 
Swin UNETR~\cite{hatamizadeh2022swin} employs Swin Transformer~(\cite{liu2021swin}) to reduce computational costs.~\cite{zhu2021region} also use SegFormer encoder~(\cite{xie2021segformer}) as the backbone, adding region prior information for more accurate breast ultrasound tumor segmentation.

\subsubsection{CNN encoder + Pure Transformer decoder}
There are limited papers involving Transformers only in Decoder as (f) in Fig~\ref{Transformer-archit in segmentation},~\cite{li2021medical} apply a pre-trained ResNet encoder to extract features and propose a squeeze-and-excitation Transformer in the decoder to learn the attention matrix. And~\cite{gong2022convtransseg} propose a PVT-style decoder to interconnect multi-resolution CNN encoder seamlessly and evaluate methods on various datasets, including ISIC2018~(\cite{tschandl2018ham10000}) and CVC-ClinicDB~(\cite{bernal2015wm}). \cite{li2022more} proposes a window attention-up-sampling with Transformer to connect different resolution features through self-attention. 

\subsubsection{Transformer encoder + Transformer decoder}
% As mentioned above, the computational cost is the major obstacle to applications of Transformer in the field of medical images. 
% Thus, as Swin Transformer~\cite{liu2021swin} has been popular for its low computational complexity, 
Recent works try to introduce Unet-like Pure Transformer, in which Transformer blocks are the basic blocks in both encoder and decoder parts (in Fig~\ref{Transformer-archit in segmentation} (g)).~\cite{sagar2021vitbis} adopt the UNet-like Transformer encoder-decoder architecture, and the Transformer input is a multi-scale feature encoded by a three-branch convolutional block with different kernel sizes.~\cite{cao2021swin} and~\cite{wu2021hepatic} use Swin Transformer as the basic block of U-Net-like networks. The latter proposes the 3D Swin Transformer and introduces inductive bias, namely region prior. 
Inspired by the shifted window design in the Swin Transformer,~\cite{peiris2021volumetric} propose a hierarchical Transformer encoder-decoder network. Especially they design a decoder block that enables parallel window-based self- and cross-attention to capture details for boundary refinement. A convex combination approach and Fourier position encoding are also added to inject complementary information.~\cite{wu2022d} design a dilated Transformer as the basic block, which conducts self-attention for pair-wise patch relations captured alternatively in both local and global scopes.~\cite{li2021gt} utilize the group Transformer ($i.e.$, a grouping architecture) to reduce computational costs and design a shape-sensitive Fourier Descriptor loss function for tooth root segmentation.~\cite{zhou2021nnformer} propose nnFormer, which jointly uses local- and global-volume-based attention at different layers to construct feature pyramids. Skip attention replaces the traditional concatenation operation in the skip connection to improve the results.

Few works also consider merging multi-scale information in medical image segmentation.~\cite{lin2021ds} improve the Swin Transformer-based U-Net architecture by adapting dual encoder subnets under different input sizes, allowing for the extraction of coarse and fine-grained representations separately. They also incorporate a Transformer Interactive Fusion module to aggregate cues between the encoder subnets for fusing multi-scale information.~\cite{huang2021missformer} propose an Enhanced Transformer-based U-Net, incorporating an Enhanced Transformer Context Bridge to combine all-level features instead of the traditional skip connection. They also introduce an efficient self-attention mechanism for spatial reduction to reduce computational costs.

We observe that these works with pure Transformer as the encoder/decoder tries to introduce the benefits of UNet-like architecture into the Transformer and the convolutional operations are inevitable in the up-/down-sampling or the patch embedding process. However, two main challenges exist. Firstly, the computational complexity of the Transformer is high, and the stacked pyramid framework exacerbates this issue. Therefore, various efficient self-attention computation methods(\emph{e.g.},~\cite{cao2021swin,wu2021hepatic,peiris2021volumetric,zhou2021nnformer,huang2021missformer,li2021gt}) have been introduced. 
Secondly, Transformer requires a large amount of data due to its lack of inductive bias(~\cite{dosovitskiy2020image}), and labeled medical images are scarce compared to labeled natural images. 
Most methods rely on transferring learning via ImageNet pretraining. ~\cite{wu2021hepatic} has introduced inductive bias, and ~\cite{tang2022self,xie2022unimiss} have explored self-supervised pretraining strategies in the medical image domain.

\subsection{Where to use (Applications)}
The Transformer-based medical applications are commonly used in multi-organ segmentation, cardiac diagnosis, polyp detection, brain tumor segmentation, and retinal segmentation tasks. Among these tasks, the universal Transformer-based medical segmentation models are often evaluated on the multi-organ segmentation task using the BCV dataset~(\cite{landman2015miccai}), as shown in Table~\ref{Transformer on BCV}. 
Our observation shows that the hybrid encoder-CNN decoder framework performs well on 2D datasets, while the methods with Transformer encoder-Transformer decoder perform better on 3D datasets. 
Moreover, part of the universal approaches is also evaluated on the cardiac diagnosis task using the ACDC dataset~(\cite{bernard2018deep}) in Table~\ref{ACDC}.

\begin{table*}[!t]
%\scriptsize
\renewcommand\arraystretch{1.25} %1.25表示1.25倍行高
\caption{An overview of Transformer methods for multi-organ segmentation on the BCV dataset.}
\label{Transformer on BCV}
%\vspace{-0.5cm}
%\begin{center}
\resizebox{\linewidth}{!}{
\begin{tabular}{@{}llllll@{}}
\toprule
\multirow{2}{*}{Method} & \multirow{2}{*}{Arch.} & \multirow{2}{*}{Type} & \multirow{2}{*}{Data split (Train:Val:Test)} & \multicolumn{2}{l}{Metric} \\ \cmidrule(l){5-6} 
 &  &  &  & Dice & HD \\ \midrule
TransUNet ~(\cite{chen2021transunet})  &
  \begin{tabular}[c]{@{}l@{}}Hybrid encoder + CNN decoder\end{tabular} &
  2D &
  18:NaN:12 (8 Organs Avg.) &
  0.7748 &
  31.69
   \\
Swin-UNet ~(\cite{cao2021swin}) &
  \begin{tabular}[c]{@{}l@{}}Transformer encoder + Transformer decoder\end{tabular} &
  2D &
  18:NaN:12 (8 Organs Avg.) &
  0.7913 &
  21.55
   \\
LeViT-UNet ~(\cite{xu2021levit}) &
  \begin{tabular}[c]{@{}l@{}}Hybrid encoder + CNN decoder\end{tabular} &
  2D &
  18:NaN:12 (8 Organs Avg.) &
  0.7853 &
  16.84
   \\
MISSFormer ~(\cite{huang2021missformer})  &
  \begin{tabular}[c]{@{}l@{}}Transformer encoder + Transformer decoder\end{tabular} &
  2D &
  18:NaN:12 (8 Organs Avg.) &
  0.8196 &
  18.2
   \\
TransClaw U-Net ~(\cite{chang2021transclaw}) &
  \begin{tabular}[c]{@{}l@{}}Hybrid encoder + CNN decoder\end{tabular} &
  2D &
  18:NaN:12 (8 Organs Avg.)&
  0.7809 &
  26.38
   \\
LiteTrans ~(\cite{xu2021litetrans}) &
  \begin{tabular}[c]{@{}l@{}}Transformer encoder + Transformer decoder\end{tabular} &
  2D &
  18:NaN:12 (8 Organs Avg.)&
  0.7791 &
  29.01
   \\
ViTBIS ~(\cite{sagar2021vitbis}) &
  \begin{tabular}[c]{@{}l@{}}Transformer encoder + Transformer decoder\end{tabular} &
  2D &
  20:NaN:10 (8 Organs Avg.)&
  0.8045 &
  21.24
   \\
ViTBIS ~(\cite{wang2022mixed}) &
  \begin{tabular}[c]{@{}l@{}}Hybrid encoder + CNN decoder\end{tabular} &
  2D &
  18:NaN:12 (8 Organs Avg.)&
  0.7859 &
  26.59
   \\
CA-GANformer ~(\cite{you2022class}) &
  \begin{tabular}[c]{@{}l@{}}Hybrid encoder + CNN decoder\end{tabular} &
  2D &
  18:NaN:12 (8 Organs Avg.) &
  0.8255 &
  22.73
   \\
DSTUNet ~(\cite{cai2022dstunet}) &
  \begin{tabular}[c]{@{}l@{}}Hybrid encoder + CNN decoder\end{tabular} &
  2D &
  18:NaN:12 (8 Organs Avg.) &
  0.8244 &
  17.83
   \\
CS-Unet ~(\cite{liu2022optimizing})  &
  \begin{tabular}[c]{@{}l@{}}Transformer encoder + Transformer decoder\end{tabular} &
  2D &
  18:NaN:12 (8 Organs Avg.) &
  0.8221 &
  27.02
   \\
DAE-Former ~(\cite{azad2022dae}) &
  \begin{tabular}[c]{@{}l@{}}Transformer encoder + Transformer decoder\end{tabular} &
  2D &
  18:NaN:12 (8 Organs Avg.) &
  0.8243 &
  17.46
   \\
TransCeption ~(\cite{azad2023enhancing}) &
  \begin{tabular}[c]{@{}l@{}}Transformer encoder + Transformer decoder\end{tabular} &
  2D &
  18:NaN:12 (8 Organs Avg.) &
  0.8224 &
  20.89
   \\
FCT ~(\cite{tragakis2023fully}) &
  \begin{tabular}[c]{@{}l@{}}Transformer encoder + Transformer decoder\end{tabular} &
  2D &
  18:NaN:12 (8 Organs Avg.) &
  0.8353 &
  NaN
   \\
Cascaded MERIT ~(\cite{rahman2023multi}) &
  \begin{tabular}[c]{@{}l@{}}Hybrid encoder + CNN decoder\end{tabular} &
  2D &
  18:NaN:12 (8 Organs Avg.) &
  0.8490 &
  13.22
   \\
CoTr ~(\cite{xie2021cotr}) &
  \begin{tabular}[c]{@{}l@{}}Hybrid encoder + CNN decoder\end{tabular} &
  3D &
  21:NaN:9 (13 Organs Avg.)&
  0.8500 &
  4.01
   \\
UNETR ~(\cite{hatamizadeh2022unetr}) &
  \begin{tabular}[c]{@{}l@{}}Pure Transformer encoder + CNN decoder\end{tabular} &
  3D &
  \begin{tabular}[c]{@{}l@{}}30 (Standard) / 80 (Free): \\ NaN:BCV test set \end{tabular} &
  \begin{tabular}[c]{@{}l@{}}0.8560 (Standard), \\ 0.8910 (Free)\end{tabular} & NaN
   \\
AFTer-UNet ~(\cite{yan2022after}) &
  \begin{tabular}[c]{@{}l@{}}Hybrid encoder + CNN decoder\end{tabular} &
  3D &
  18:NaN:12 (8 Organs Avg.)&
  0.8102  & NaN
   \\
UTNetV2 ~(\cite{gao2022multi}) &
  \begin{tabular}[c]{@{}l@{}}Transformer encoder + Transformer Decoder\end{tabular} &
  3D &
  80\%:20\%:NaN (13 Organs Avg.)&
  0.8514 &
  15.78
   \\
D-Former ~(\cite{wu2022d}) &
  \begin{tabular}[c]{@{}l@{}}Transformer encoder + Transformer decoder\end{tabular} &
  3D &
  18:NaN:12 (8 Organs Avg.)&
  0.8883  & NaN
   \\
nnFormer ~(\cite{zhou2021nnformer}) &
  \begin{tabular}[c]{@{}l@{}}Transformer encoder + Transformer decoder\end{tabular} &
  3D &
  18:NaN:12 (8 Organs Avg.)&
  0.8657 &
  10.63
   \\
FINE ~(\cite{themyr2022memory}) &
  \begin{tabular}[c]{@{}l@{}}Transformer encoder + Transformer decoder\end{tabular} &
  3D &
  18:NaN:12 (7 Organs Avg.)&
  0.8710 &
  9.2
   \\
UNETR++ ~(\cite{shaker2022unetr++}) &
  \begin{tabular}[c]{@{}l@{}}Transformer encoder + Transformer decoder\end{tabular} &
  3D &
  18:NaN:12 (8 Organs Avg.)&
  0.8722 &
  7.53
   \\
 \bottomrule
\end{tabular}
}
\end{table*}

\begin{table*}[!h]
\scriptsize
\renewcommand\arraystretch{1.25} %1.25表示1.25倍行高
\caption{An overview of Transformer methods on cardiac segmentation on the ACDC dataset, in which RV is the right ventricle, LV means the left ventricle and the Myo is the myocardium. Ave. is the average dice.}
\label{ACDC}
%\vspace{-0.5cm}
%\begin{center}
\resizebox{\linewidth}{!}{
\begin{tabular}{llllllll}
\hline
\multicolumn{1}{c}{\multirow{2}{*}{Method}}                          & \multicolumn{1}{c}{\multirow{2}{*}{Arch.}} & \multicolumn{1}{c}{\multirow{2}{*}{Type}} & \multicolumn{1}{c}{\multirow{2}{*}{Data split(Train: Val:   Test)}} & \multicolumn{4}{c}{Metric}        \\ \cline{5-8} 
\multicolumn{1}{c}{}                                                 & \multicolumn{1}{c}{}                       & \multicolumn{1}{c}{}                      & \multicolumn{1}{c}{}                                                & RV     & Myo    & LV     & Ave.   \\ \hline
TransUNet ~(\cite{chen2021transunet})       & Hybrid encoder + CNN decoder               & 2D                                        & 70: 10: 20                                                          & 0.8886 & 0.8453 & 0.9573 & 0.8971 \\
Swin-UNet ~(\cite{cao2021swin})             & Transformer encoder + Transformer decoder  & 2D                                        & 70: 10: 20                                                          & 0.8855 & 0.8562 & 0.9583 & 0.9000 \\
LeViT-UNet ~(\cite{xu2021levit})              & Hybrid encoder + CNN decoder               & 2D                                        & 80: NaN: 20                                                         & 0.8955 & 0.8764 & 0.9376 & 0.9032 \\
MISSFormer ~(\cite{huang2021missformer})   & Transformer encoder + Transformer decoder  & 2D                                        & 70: 10: 20                                                          & 0.8955 & 0.8804 & 0.9499 & 0.9086 \\
LiteTrans ~(\cite{xu2021litetrans})          & Transformer encoder + Transformer decoder  & 2D                                        & 80: 20: NaN                                                         & 0.8966 & 0.8797 & 0.8533 & 0.8966 \\
DSTUNet ~(\cite{cai2022dstunet})           & Hybrid encoder + CNN decoder               & 2D                                        & 70: 10: 20                                                          & 0.8036 & 0.8177 & 0.8834 & 0.8350 \\
CS-Unet ~(\cite{liu2022optimizing})         & Transformer encoder + Transformer decoder  & 2D                                        & 70: 10: 20                                                          & 0.8920 & 0.8947 & 0.9542 & 0.9137 \\
FCT ~(\cite{tragakis2023fully})         & Transformer encoder + Transformer decoder  & 2D                                        & 70: 10: 20                                                          & 0.9264 & 0.9051 & 0.9550 & 0.9302 \\
Cascaded MERIT ~(\cite{rahman2023multi})  & Hybrid encoder + CNN decoder               & 2D                                        & 70: 10: 20                                                          & 0.9023 & 0.8953 & 0.9580 & 0.9185 \\
D-Former ~(\cite{wu2022d})                   & Transformer encoder + Transformer decoder  & 3D                                        & 70: 10: 20                                                          & 0.9133 & 0.8960 & 0.9593 & 0.9229 \\
nnFormer ~(\cite{zhou2021nnformer})         & Transformer encoder + Transformer decoder  & 3D                                        & 70: 10: 20                                                          & 0.9094 & 0.8958 & 0.9565 & 0.9206 \\
UNETR++ ~(\cite{shaker2022unetr++})       & Transformer encoder + Transformer decoder  & 3D                                        & 70: 10: 20                                                          & 0.9189 & 0.9061 & 0.9600 & 0.9283 \\ \hline
\end{tabular}
}
\end{table*}

\textbf{Brain Tumor.}
Automated and accurate segmentation of brain tumors plays an essential role in the timely diagnosis of neurological diseases, and Transformer-based methods have been proposed for these tasks effectively.~\cite{jun2021medical} firstly introduce Transformer into the brain tumor segmentation task with a ViT-style pure Transformer encoder. Later, TransBTS~(\cite{wang2021transbts}) is proposed to take advantage of both CNN and Transformer in the local and global feature extraction, which inserts a Transformer module in the bottleneck of U-Net framework and inspires some following works on improving it(\emph{e.g.},~\cite{jia2021bitr,dobko2021combining,li2022transbtsv2,hatamizadeh2022swin}). 
To further apply Transformer at each stage,~\cite{liang2022transconver} combine the convolution and Transformer blocks in parallel and integrate the two feature maps by a cross-attention fusion as an encoder basic block. Moreover, Swin Transformer-based encoder architectures~(\cite{liang20223d}) and Swin-Transformer based encoder-decoder architectures~(\cite{jiang2022swinbts,liang2022btswin,peiris2022hybrid}) are designed for brain tumor segmentation to reduce the computational costs. 
It is noticed that multi-modality fusion is essential for precise brain tumor segmentation from MRI, thus~\cite{li2022transiam,zhang2022mmformer} and \cite{xing2022nestedformer} adopt multi-branch encoder for different modalities separately and the Transformer module act as the bottleneck between the encoder and the decoder. The proposed Transformer module consists of a self-attention module to enhance long-term dependencies within individual modalities and a cross-attention block to catch cross-modality contextual information. These methods are all experimented on the BraTS dataset in Table~\ref{trans-brain}.

\begin{table*}[!h]
\scriptsize
\renewcommand\arraystretch{1.25} %1.25表示1.25倍行高
\caption{An overview of Transformer methods on brain tumor segmentation application. The results are all tested on the BraTs validation dataset ~(\cite{menze2014multimodal}), in which WT is the whole tumor, TC means the tumor core and ET is the enhanced tumor.}
\label{trans-brain}
%\vspace{-0.5cm}
%\begin{center}
\resizebox{\linewidth}{!}{
\begin{tabular}{@{}cccccccl@{}}
\toprule
\multirow{2}{*}{Method}                                                             & \multirow{2}{*}{Arch.}                         & \multirow{2}{*}{Dataset} & \multicolumn{3}{c}{Metric}  & \multirow{2}{*}{Data   split(Train:Val:Test)} & \multicolumn{1}{c}{\multirow{2}{*}{Modality(Type)}} \\
                                                                                    &                                                &                          & Dice WT & Dice TC & Dice ET &                                               & \multicolumn{1}{c}{}                                \\ \midrule
\cite{zhang2022mmformer}                   & Hybrid encoder + CNN decoder                                & BraTs 2018               & 0.8964  & 0.8578  & 0.7761  & 285:66:NaN                                    & MRI (3D)                                            \\ \cmidrule(l){3-8} 
\multirow{2}{*}{\cite{liang2022btswin}}          & \multirow{2}{*}{Transformer encoder + Transformer decoder}       & BraTs 2018               & 0.9174  & 0.8553  & 0.8193  & 285:66:NaN                                    & MRI (3D)                                            \\ \cdashline{3-8}[0.8pt/5pt]
                                                                                    &                                                & BraTs 2019               & 0.9028  & 0.8173  & 0.7838  & 335:125:NaN                                   & MRI (3D)                                            \\ \cmidrule(l){3-8} 
\multirow{2}{*}{\cite{liang2022transconver}} & \multirow{2}{*}{Hybrid encoder + CNN decoder} & BraTs2018                & 0.9157  & 0.8568  & 0.8173  & 285:66:NaN                                    & MRI (3D)                                            \\ \cdashline{3-8}[0.8pt/5pt]
                                                                                    &                                                & BraTs 2019               & 0.9019  & 0.8257  & 0.7840   & 335:125:NaN                                   & MRI (3D)                                            \\ \cmidrule(l){3-8} 
\multirow{2}{*}{\cite{wang2021transbts}}   & \multirow{2}{*}{Hybrid encoder + CNN decoder}               & BraTs 2019               & 0.9000     & 0.8194  & 0.7893  & 335:125:NaN                                   & MRI (3D)                                            \\ \cdashline{3-8}[0.8pt/5pt]
                                                                                    &                                                & BraTs 2020               & 0.9009  & 0.8173  & 0.7873  & 369:125:NaN                                   & MRI (3D)                                            \\ \cmidrule(l){3-8} 
\cite{jun2021medical}                       & Hybrid encoder + CNN decoder                        & BraTs 2020               & 0.8695  & 0.6363  & 0.5063  & 369:125:NaN                                   & MRI (3D)                                            \\ \cmidrule(l){3-8} 
\cite{xing2022nestedformer}                      & Pure Transformer encoder + CNN decoder                              & BraTs 2020               & 0.9200    & 0.8640   & 0.8000     & 315:17:37                                     & MRI (3D)                                            \\ \cmidrule(l){3-8} 
\multirow{2}{*}{\cite{jiang2022swinbts}}  & \multirow{2}{*}{Hybrid encoder + CNN decoder}       & BraTs 2020               & 0.8906  & 0.8030   & 0.7736  & 369:125:NaN                                   & MRI (3D)                                            \\ \cdashline{3-8}[0.8pt/5pt]
                                                                                    &                                                & BraTs 2021               & 0.9183  & 0.8475  & 0.8321  & 1251:219:NaN                                  & MRI (3D)                                            \\ \cmidrule(l){3-8} 
\multirow{2}{*}{\cite{liang20223d}} & \multirow{2}{*}{Pure Transformer encoder + CNN decoder}       & BraTs 2020               & 0.9076  & 0.8420   & 0.7948  & 371:127:NaN                                   & MRI (3D)                                            \\ \cdashline{3-8}[0.8pt/5pt]
                                                                                    &                                                & BraTs 2021               & 0.9183  & 0.8475  & 0.8321  & 1251:219:NaN                                  & MRI (3D)                                            \\ \cmidrule(l){3-8} 
\cite{jia2021bitr}                          & Hybrid encoder + CNN decoder                               & BraTs 2021               & 0.9097  & 0.8434  & 0.8187  & 1251:219:NaN                                  & MRI (3D)                                            \\ \bottomrule
\end{tabular}
}
\end{table*}

\textbf{Polyp.}
Accurate polyp segmentation is a challenge due to the variable size and shape of polyps, as well as the indistinct boundaries between polyps and mucosa. Hence, local information is essential and the Transformer block is always combined with the convolutional block in the polyp segmentation task (as shown in Table~\ref{trans-polyp}).~\cite{tomar2022transresu} combine a Swin Transformer block and a dilated convolutional block as a basic module at the bottleneck of U-Net. \cite{dong2021polyp, wang2022stepwise, sanderson2022fcn, park2022swine}, and \cite{duc2022colonformer} introduce PVT-like network to the polyp segmentation task. It is noted that a progressive decoder is also proposed for improved local emphasis and stepwise feature aggregation in works(\emph{e.g.,}~\cite{wang2022stepwise, sanderson2022fcn}).

\begin{table*}[!t]
%\scriptsize
\renewcommand\arraystretch{1.25} %1.25表示1.25倍行高
\caption{An overview of Transformer methods on polyp segmentation application}
\label{trans-polyp}
%\vspace{-0.5cm}
%\begin{center}
\resizebox{\linewidth}{!}{
\begin{tabular}{@{}cccccc@{}}
\toprule
Method                                                     & Arch.                                          & Dataset      & Metric                    & Data split(Train:Test) & Modality(Type) \\ \midrule
\multirow{5}{*}{\cite{dong2021polyp}}      & \multirow{5}{*}{Pure Transformer encoder + CNN decoder}       & 1.Kvasir-SEG ~(\cite{jha2020kvasir}) & Dice: 0.9170 mIoU: 0.8640 & 900:100                & Endoscopy (2D) \\ \cdashline{3-6}[0.8pt/5pt]
                                                           &                                                & 2.ClinicDB ~(\cite{bernal2015wm})   & Dice: 0.9370 mIoU: 0.8890 & 548:64                 & Endoscopy (2D) \\ \cdashline{3-6}[0.8pt/5pt]
                                                           &                                                & 3.ColonDB ~(\cite{tajbakhsh2015automated})    & Dice: 0.8080 mIoU: 0.7270 & NaN:380                & Endoscopy (2D) \\ \cdashline{3-6}[0.8pt/5pt]
                                                           &                                                & 4.Endoscene ~(\cite{vazquez2017benchmark})  & Dice: 0.9000 mIoU: 0.8330 & NaN:60                 & Endoscopy (2D) \\ \cdashline{3-6}[0.8pt/5pt]
                                                           &                                                & 5.ETIS ~(\cite{silva2014toward})       & Dice: 0.7870 mIoU: 0.7060 & NaN;196                & Endoscopy (2D) \\ \cmidrule(l){3-6} 
\multirow{2}{*}{\cite{wang2022stepwise}}       & \multirow{2}{*}{Pure Transformer encoder + CNN decoder}       & 1.ClinicDB ~(\cite{bernal2015wm})   & Dice: 0.9447 mIoU: 0.8995 & 548:64                 & Endoscopy (2D) \\ \cdashline{3-6}[0.8pt/5pt]
                                                           &                                                & 2.Kvasir-SEG ~(\cite{jha2020kvasir}) & Dice: 0.9357 mIoU: 0.8905 & 900:100                & Endoscopy (2D) \\ \cmidrule(l){3-6} 
\multirow{2}{*}{\cite{sanderson2022fcn}} & \multirow{2}{*}{Hybrid encoder + CNN decoder} & 1.Kvasir-SEG ~(\cite{jha2020kvasir}) & Dice: 0.9385 mIoU: 0.8903 & 900:100                & Endoscopy (2D) \\ \cdashline{3-6}[0.8pt/5pt]
                                                           &                                                & 2.ClinicDB ~(\cite{bernal2015wm})   & Dice: 0.9469 mIoU: 0.9020 & 548:64                 & Endoscopy (2D) \\ \cmidrule(l){3-6} 
\multirow{5}{*}{\cite{park2022swine}}      & \multirow{5}{*}{Hybrid encoder + CNN decoder} & 1.Kvasir-SEG ~(\cite{jha2020kvasir}) & Dice: 0.9200 mIoU: 0.8700 & 900:100                & Endoscopy (2D) \\ \cdashline{3-6}[0.8pt/5pt]
                                                           &                                                & 2.ClinicDB ~(\cite{bernal2015wm})   & Dice: 0.9380 mIoU: 0.8920 & 550:62                 & Endoscopy (2D) \\ \cdashline{3-6}[0.8pt/5pt]
                                                           &                                                & 3.ColonDB ~(\cite{tajbakhsh2015automated})    & Dice: 0.8040 mIoU: 0.7250 & NaN:380                & Endoscopy (2D) \\ \cdashline{3-6}[0.8pt/5pt]
                                                           &                                                & 4.Endoscene ~(\cite{vazquez2017benchmark})  & Dice: 0.7580 mIoU: 0.6870 & NaN:60                 & Endoscopy (2D) \\ \cdashline{3-6}[0.8pt/5pt]
                                                           &                                                & 5.ETIS ~(\cite{silva2014toward})       & Dice: 0.9060 mIoU: 0.8420 & NaN;196                & Endoscopy (2D) \\ \cmidrule(l){3-6} 
\multirow{5}{*}{\cite{duc2022colonformer}}       & \multirow{5}{*}{Pure Transformer encoder + CNN decoder
}       & 1.Kvasir-SEG~(\cite{jha2020kvasir}) & Dice: 0.9240 mIoU: 0.8760 & 900:100                & Endoscopy (2D) \\ \cdashline{3-6}[0.8pt/5pt]
                                                           &                                                & 2.ClinicDB ~(\cite{bernal2015wm})   & Dice: 0.9320 mIoU: 0.8840 & 548:64                 & Endoscopy (2D) \\ \cdashline{3-6}[0.8pt/5pt]
                                                           &                                                & 3.ColonDB ~(\cite{tajbakhsh2015automated})    & Dice: 0.8110 mIoU: 0.7330 & NaN:380                & Endoscopy (2D) \\ \cdashline{3-6}[0.8pt/5pt]
                                                           &                                                & 4.Endoscene ~(\cite{vazquez2017benchmark})      & Dice: 0.9060 mIoU: 0.8420 & NaN:60                 & Endoscopy (2D) \\ \cdashline{3-6}[0.8pt/5pt]
                                                           &                                                & 5.ETIS ~(\cite{silva2014toward})       & Dice: 0.8010 mIoU: 0.7220 & NaN;196                & Endoscopy (2D) \\ \cmidrule(l){3-6} 
\multirow{2}{*}{\cite{tomar2022transresu}}     & \multirow{2}{*}{Hybrid encoder + CNN decoder}               & 1.Kvasir-SEG ~(\cite{jha2020kvasir}) & Dice: 0.8884 mIoU: 0.8214 & 880:120                & Endoscopy (2D) \\ \cdashline{3-6}[0.8pt/5pt]
                                                           &                                                & 2.BKAI-IGH ~(\cite{ngoc2021neounet})   & Dice: 0.9154 mIoU: 0.8568 & 80:10:10               & Endoscopy (2D) \\ \bottomrule
\end{tabular}
}
\end{table*}

\textbf{Retinal.}
The applications of transformer in eye image include fovea localization~(\cite{song2022bilateral}), iris segmentation~(\cite{wei2021toward}), retinal vessel segmentation~(\cite{chen2022pcat, jiang2022mtpa_unet}), and retinopathy lesion segmentation~(\cite{huang2022rtnet}). The network architecture is various and each of them makes some special designs.~\cite{song2022bilateral} propose a bi-branch CNN encoder, which consists of a main branch for retinal images and a vessel branch for pre-segmentation of vessel images. A Transformer block is used as the bottleneck in the main branch, similar to TransUnet~(\cite{chen2021transunet}), to remove vessel interference and improve fovea localization.~\cite{wei2021toward} design a bilateral self-attention module to apply spatial and visual branches to learn contextual clues for two characteristics in the iris segmentation and build a Transformer encoder-decoder architecture. 
While~\cite{huang2022rtnet} also propose a relation Transformer block to catch relationships between the lesions and vessel features. 
Moreover,~\cite{chen2022pcat} propose a patch convolution attention Transformer in an encoder-decoder framework and~\cite{jiang2022mtpa_unet} design modified multi-head attention in the Transformer block and combine the CNN and Transformer block at each layer in the encoder for vessel retinal segmentation, both of them attempt to take the advantages of CNN and Transformers.

\textbf{Conclusion.}
The Transformer-based medical segmentation methods can be categorized into different categories based on the encoder-decoder network components, $i.e.$, hybrid encoder + CNN decoder, Pure Transformer encoder + CNN decoder, CNN encoder + Pure Transformer decoder, Transformer encoder + Transformer decoder, and further subdivide them on the basis of inserted locations. 
The Transformer can be inserted into the CNN network as a plug-in, either as a bottleneck or multiple layers, to model long-term dependencies in various scales while retaining accurate spatial information and inductive bias from CNN. However, the interleaved CNN-Transformer structure can destroy consistency and delivery between features, which may be addressed by adopting a UNet-like Transformer architecture.
Besides, we also discuss the performance of these methods for several important applications on popular benchmarks. 
However, it is important to note that the performance of these Transformer-based methods varies depending on the specific medical application and the benchmark dataset used for evaluation. For example, some methods may perform better on multi-organ segmentation tasks while others may excel at cardiac diagnosis. Therefore, a thorough and impartial benchmark and a uniform public dataset are necessary to evaluate and compare the complexity and performance of these models across different medical applications. This would provide a clearer understanding of the strengths and weaknesses of each method and allow for more informed choices in selecting the most appropriate approach for a specific medical segmentation task.

% When Transformer is integrated into the CNN network as a plug-in, a direct way is inserted Transformer as the bottleneck. However, this kind of method only models long-term dependencies on the top-level features. Thus Transformer is added in multi-layers (in the skip connection or after the CNN layers) in the CNN framework to catch long-range relationships in various scales while retaining the accurate spatial information and inductive bias from CNN. And the parallel CNN-Transformer structure attempts to take full advantage of complementary local-global information at the cost of space. Moreover, though the convolutional layer supplements some local context and get better performance, the interleaved CNN-Transformer structure destroys the consistency and delivery between features. 
% Thus some work on adopting a UNet-like Transformer based on the advent of an efficient Transformer. We also list a table (Table\ref{Transformer on BCV}) of those methods on the multi-atlas labeling beyond the cranial vault--workshop and challenge (BCV) dataset~\cite{landman2015miccai} and we observe that hybrid encoder-CNN decoder framework have competitive results on the 2D dataset while the methods with Transformer encoder-Transformer decoder performs better on the 3D dataset. However, a thorough and impartial benchmark and a uniform public dataset need to be established to judge the complexity and performance of these models. 

\section{Discussion}
\label{sec5}
% In this section, we will provide a brief overview of the evolution of the attention mechanism in medical image segmentation. Additionally, we will analyze the distinctions between Non-Transformer methods and Transformer-based methods. 

\subsection{Summary}
\begin{figure*}
    \centering
    \centerline{\includegraphics[width=0.8\linewidth]{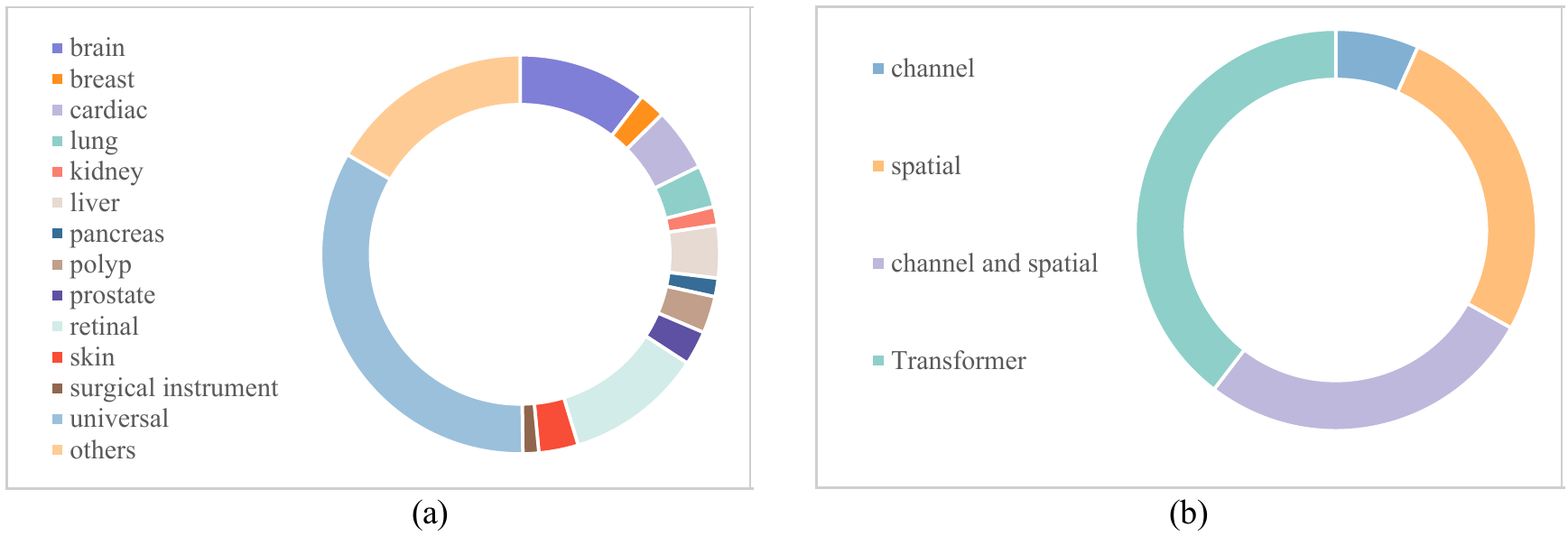}}
    \caption{The charts present statistics on attention-based medical image segmentation methods according to their applications and attention types. (a) displays the distribution of task-specific applications. (b) shows the ratio of different attention types.}
    \label{distribution}
\end{figure*}

We have reviewed the diverse applications of the attention mechanism in medical image segmentation. The attention mechanism has been widely used in various tasks such as brain segmentation, breast segmentation, cardiac segmentation, lung segmentation, kidney segmentation, liver segmentation, pancreas segmentation, polyp segmentation, prostate segmentation, retinal segmentation, skin segmentation, and other miscellaneous tasks, as shown in Fig~\ref{distribution} (a). Most attention-based methods are designed to be universal and can be applied to different tasks. Various attention mechanisms have been introduced into medical image segmentation, as depicted in Fig~\ref{distribution} (b). This has contributed to the steady and sustained growth of attention-based methods in the field.

The traditional attention mechanism, which mimics the human visual system, focuses on the most important regions of an image and discards irrelevant parts, shifting the machine learning paradigm from large-scale vector transformation to more conscious processes. This approach is computationally efficient, low-cost, and interpretable, which is critical for medical images due to their large image resolution and small data scale. In the segmentation domain, channel and spatial attention are the most common non-Transformer attention types. Channel attention adaptively reweights each channel, acting as a feature selection process focusing on the target object (tissues, lesions, or organs). Spatial attention, conversely, can be seen as a spatial region selection mechanism that focuses on the area of interest or abnormality. We observe that spatial attention is more popular than channel attention, likely because edge information is significant for blurring boundaries in tasks like polyp segmentation, prostate segmentation, and skin lesion segmentation, among others. 
The non-Transformer attention mechanism is usually combined with convolutional layers in the encoder, decoder, or skip connection.  

The limited receptive field and stationary weights of the traditional attention mechanisms have prompted the rapid and widespread adoption of the Transformer in numerous vision tasks, due to its ability to model long-range dependencies. As the Transformer was successfully applied in the field of vision, it has also gradually been applied in a very simple way to the field of medical image segmentation, such as being incorporated at the bottom of UNet or completely replacing standard convolutions. However, convolutional operations can capture local information, which is a weakness of Transformers. Consequently, researchers have improved Transformer blocks by combining them with convolutional blocks in multi-layers in a serial or parallel fashion to take advantage of multi-resolution features for more comprehensive modeling. Moreover, to address time and computational cost issues, DETR~(\cite{zhu2020deformable}), Swin Transformer~(\cite{liu2021swin}), and LeViT~(\cite{graham2021levit}) have replaced the vanilla Transformer block with more efficient ones. Additionally, researchers have continued to develop the integration of the Transformer into U-Net style architectures (e.g., pure Transformer encoder with CNN decoder) to leverage the U-like architecture's benefits.

\subsection{Future Challenge}
%需要建立更加标准的数据集和评估平台（例如attention加在每个位置的影响力，不同结构的Transformer应该怎么对比谁优谁劣）
% Despite the exciting application of attention-based medical segmentation, there remain several open future challenges.
Despite the numerous successful applications of attention mechanisms in medical image segmentation, several challenges still need to be addressed.

\subsubsection{Task-specific Attention}
% After the literature review on attention-based medical image segmentation, it is a pity that most methods, including the traditional attention mechanism and the Transformer-based model, are not designed for segmentation-specific. Though spatial attention is well-suited to dense prediction tasks such as semantic and object detection due to its capability as an adaptive selection of areas of interest, some attention mechanisms may be introduced directly from original tasks without any change. And Transformer is firstly applied as original ViT in the medical image segmentation, which means more task-specific architectures are in need to design such as an auxiliary branch, task-specific attention, etc.
Different tasks may require distinct levels of attention to specific features or areas of interest. For instance, in lung nodule segmentation, a multi-scale attention mechanism is crucial for handling lesions of varying sizes. In brain tumor segmentation, a unique attention mechanism is essential for differentiating between tumor and regular tissue. The required attention mechanism depends on the task at hand. Unfortunately, after reviewing the literature in this field, it has been found that most existing methods are not specifically designed for specialized medical image segmentation tasks. As a result, exploring this area further would be incredibly valuable.

\subsubsection{Robustness}
Although attention mechanisms can provide insight into the information that the model is focusing on, there may be instances where the attention results are partially or completely incorrect, leading to inaccurate model predictions. Currently, there is limited research focusing on the failure cases of attention models in medical image segmentation. It is crucial to study these cases in depth to understand the limitations and potential drawbacks of attention mechanisms, as well as to improve their robustness and generalizability in practical applications. By analyzing and addressing these failure cases, researchers can develop more reliable and accurate attention-based models, which in turn will enhance their adoption in clinical settings and their impact on patient care.

\subsubsection{Standard Evaluation}
The varying datasets, image processing modes, and data partitioning used for evaluation in the literature make it challenging to compare the accuracy and validity of the surveyed methods. It is essential to establish standard and diverse datasets that can fully represent the diversity of medical images, as well as a standardized training and validation process to confirm the effectiveness of proposed models. Additionally, exploring the performance impact of different types of attention mechanisms in various network locations or architectural designs can provide insights into better ways to apply attention mechanisms. This will help researchers identify optimal strategies for incorporating attention mechanisms into medical image segmentation tasks, ultimately leading to more accurate and efficient models.

\subsubsection{Multi-modality \& Multi-task}
Multi-task learning helps improve a model's generalizability and performance by leveraging the relevance between different tasks. Thus, it is meaningful to build models capable of handling multiple tasks.~\cite{park2021federated} propose a federated split vision Transformer for COVID-19 diagnosis by co-training the segmentation, classification, and detection tasks. Beyond that, different medical imaging modalities provide complementary properties for diagnosis.~\cite{song2021deep} utilize the attention mechanism to model the pairwise relation between Optimal Coherence Tomography features and Visual Field features, in which the complementary information is passed from one modality to another by utilizing the Transformer model. While~\cite{xie2022unimiss} propose a Pyramid Transformer U-Net network to learn representations from diverse dimension data and transfer features to various downstream tasks through the switchable patch embedding layers, which outperforms the advanced SSL counterparts substantially. Thus, it is expected to see more work combing multi-task and multi-modality as the Transformer is inherently suitable for various sequence-based inputs.

\subsubsection{Complexity}
The application of Transformers in medical image segmentation is often hindered by the high computational and memory costs. As a result, researchers have introduced efficient Transformer blocks into models. However, the multi-level feature fusion module, which can integrate features both locally and globally to enhance segmentation performance and address data sparsity issues, remains a high-cost component. CoTr~(\cite{xie2021cotr}) introduces the deformable self-attention mechanism to tackle this challenge. We anticipate further research and solutions in this specific direction, as it is essential for making Transformer-based architectures more accessible and efficient in medical image segmentation tasks.

\section{Conclusion}
\label{sec6}

In this paper, we have reviewed over 300 articles related to attention-based medical image segmentation applications, systematically surveying and summarizing the literature grouped into Non-Transformer and Transformer categories. We have provided an in-depth view of recent trends and future challenges in this field. Our aim is to give researchers a deeper understanding of the attention mechanisms applied in medical image segmentation and to serve as a springboard for future research. By examining the current state of the art and identifying potential areas for improvement, we hope to inspire the development of more advanced and effective attention-based techniques for medical image segmentation, ultimately contributing to better diagnosis and treatment in healthcare.

{
        \bibliographystyle{model2-names.bst}\biboptions{authoryear}
	\bibliography{refs}
}

\end{document}